\documentclass[twocolumn,epjc3]{svjour3} 

\RequirePackage[english]{babel}
\RequirePackage[T1]{fontenc} 
\RequirePackage[utf8]{inputenc}
\RequirePackage{amsmath}
\RequirePackage{amsfonts}
\RequirePackage{amssymb}
\RequirePackage{mathptmx}

\RequirePackage{graphicx}
\RequirePackage{subfigure}
\RequirePackage{color}

\RequirePackage[colorlinks=true,pdftoolbar = true, pdfstartview = FitH, pdfmenubar = true]{hyperref}

\newcommand{\weff}{\mathrm{w}_{\mathrm{eff}}}

\journalname{Eur. Phys. J. C}
\begin{document}

\title{Global Portraits of Nonminimal Inflation}

\author{Laur J\"arv \thanksref{e1, addr1} \and Alexey Toporensky \thanksref{e2,addr2,addr3}}
\institute{Institute of Physics, University of Tartu, W.\ Ostwaldi 1, 50411 Tartu, Estonia \label{addr1} \and Sternberg Astronomical Institute, Lomonosov Moscow State University, Universitetsky pr.\ 13, Moscow, 119991, Russia \label{addr2} \and Kazan Federal University, Kremlevskaya str.\ 18, Kazan, 420008, Russia \label{addr3}}
\thankstext{e1}{laur.jarv@ut.ee}
\thankstext{e2}{atopor@rambler.ru}

\date{Received: 18 October 2021, revised: 15 December 2021, second revision: 30 December 2021}

\maketitle

\begin{abstract}
We reconsider the dynamical systems approach to analyze inflationary universe in the Jordan frame models of scalar field  nonminimally coupled to curvature. The adopted set of variables allows us to clearly distinguish between different asymptotic states in the phase space,  including the kinetic and inflationary regimes. Inflation is realized as a heteroclinic trajectory originating either at infinity from a nonhyperbolic asymptotic de Sitter point or from a regular saddle de Sitter point. We also present a comprehensive picture of possible initial conditions leading to sufficient inflationary expansion and show their extent on the phase diagrams. In addition we comment on the slow roll conditions applicable in the Jordan frame and show how they approximate the leading inflationary ``attractor solution''. As particular examples we portrait quadratic and quartic potential models and note that  increasing the nonminimal coupling diminishes the range of good initial conditions in the quadratic case, but enlarges is in the quartic case. 
\end{abstract}


\section{Introduction}

Phase diagrams have been used intensively for a large variety of dynamical systems as they give a good visual tool to understand the global features of the system. Isotropic cosmological evolution with a scalar field has a particular advantage to be presented via such diagrams, especially in a spatially flat case, where the number of independent degrees of freedom is equal to two, giving a possibility to encode all the dynamics into a planar diagram (see Refs.\ \cite{Wainwright_book,Coley:2003mj,Bahamonde:2017ize} for reviews). The phase portrait method is usually applied and is particularly effective when describing trajectories between hyperbolic fixed points.

On the other hand, the early studies of possibly the most important application of scalar field cosmology, the inflationary regime \cite{Linde:1981mu,Albrecht:1982wi}, have less often used this method, and in general found not so profound impact from the dynamical systems theory.\footnote{Let us note, however, that phase portraits appeared in the early papers on the alternative purely geometrical inflationary models \cite{Starobinsky:1980te,Gurovich:1979xg}.}
The obvious reason is that the physical requirements for this method to be useful for early Universe are rather contradictory from the viewpoint of dynamical systems. First, the inflationary regime should attract a big part of neighboring trajectories in order to be unaffected by the ``fine tuning'' problem -- we should be able to start from rather general initial conditions
at the beginning of the inflationary stage 
\cite{1978SvAL....4...82S,Belinsky:1985zd,Belinsky:1987gy,Linde:1985ub,PIRAN1985331}. That is why this regime have been frequently refered to as an ``inflationary attractor'' \cite{Liddle:1994dx}. However, it can not be an attractor in a proper sense of a fixed point, since in the  dynamical systems theory it would imply that the system stays in this state forever, which is inappropriate for obvious physical reasons. Our present world looks quite different, and we need a mechanism to exit the inflationary ``attractor.''\footnote{There have been some attempts to describe inflation as a ``quasi'' fixed point, though \cite{UrenaLopez:2007vz,Kiselev:2008zm,Kiselev:2009xm}.}

Although qualitatively noted already earlier \cite{Felder:2002jk,UrenaLopez:2011ur}, an important step in applying the dynamical systems paradigm to inflation was achieved by Alho and Uggla \cite{Alho:2014fha} who showed that for a minimally coupled massive scalar field the inflationary regime corresponds to a 1-dimensional 
center submanifold originating from an asymptotic nonhyperbolic fixed point, i.e.\ emerging from the eigendirection of zero real eigenvalue. The structure of the center manifold explains why nearby solutions are attracted to this solution, and provides a global analytical approximate description of the corresponding dynamics, confirming and improving the slow roll approximation scheme of Ref.\ \cite{Liddle:1994dx}.
A similar picture persists after including a perfect fluid \cite{Alho:2015cza} as well as for the E- and T-model $\alpha$-attractors \cite{Alho:2017opd}. In the present work we turn on a nonminimal coupling between the scalar field and curvature,\footnote{Since the inflationary dynamics of some modified gravity models is effectively equivalent to the nonminimally coupled field, e.g.\ \cite{He:2018gyf}, our results are applicable to these theories too.} and show that the same feature can be also realized by a saddle point and an unstable submanifold originating from it. In either case the ``attactor solution'' is given by a heteroclinic orbit running from a fixed point situated at a (local or asymptotic) maximum of the scalar field effective potential to another fixed point residing at the minimum. In the global phase space this trajectory acts as a separatrix between different classes of solutions.\footnote{Recently the relevance of a primordial saddle point for the description of inflation was also  pointed out by Refs.\ \cite{Quiros:2020bcg,Hrycyna:2020jmw} in different variables, while the role of separatrices was emphasized by Ref.\ \cite{Alvarez:2019vbp} in the Hamilton-Jacobi formalism.}

The ``attractor solution'' of Ref.\ \cite{Alho:2014fha} originates from an asymptotic fixed point which corresponds to infinite values of the scalar field as well as of its time derivative, and was referred to as de Sitter point. The authors of that paper remarked, however, that this de Sitter point should not be considered as a ``usual'' de Sitter solution in cosmology which implies finite values of $\phi$ and $\dot \phi$. Indeed, the asymptotic form of the ``de Sitter'' solution in infinity may be quite different from the $\phi=const$ solution from a naive point of view. For example, for the massive scalar field case the asymptotic in question is known to have a linear decay of the scalar field value, while for the quartic potential this decay is even more sharp being exponential. Sometimes in the literature, though rather rarely, these two regimes are distinguished even at the level of dynamical system analysis. One of the examples is the Starobinsky inflation in quadratic gravity \cite{1978SvAL....4...82S,Starobinsky:1983zz}. The Hubble parameter $H$ in the Jordan frame changes linearly in time and tends to infinity at the corresponding fixed point. On the other hand, a general quadratic gravity contains de Sitter solution in a proper sense, with $H=const$. In the paper of Barrow and Hervik \cite{Barrow:2006xb} one of the variables used for describing the corresponding dynamical system is chosen to be inversely proportional to $H^2$. Evidently, this variable tends to zero at Starobinsky inflation, and to non-zero constant value for a proper de Sitter solution, which enables the authors to distinguish them. Usually, however, variables used in dynamical analysis have the same values for both finite and asymptotic de Sitter solution. Taking all these precautions we will call the point at infinity as asymptotic de Sitter for brevity.

The principal task of the current paper is to present global phase portraits of the scalar field inflationary dynamics, illustrating the ``attractor solution'' and distinguishing various asymptotic regimes. To foster physical insight, we need to choose a helpful set of variables to visualise the diagrams. The set employed in Refs.\ \cite{Alho:2014fha,Rendall:2001it} is well suited from the mathematical point of view as it renders the system regular, but these variables lack direct physical interpretation. On the other hand, the set $(\tfrac{\dot{\phi}}{H},\tfrac{\sqrt{V}}{H})$ \cite{Copeland:1997et} is nicely related to the kinetic and potential energy densities of the scalar field in the Friedmann equation, and thus favored in a number of studies \cite{Szydlowski:2008in,Szydlowski:2013sma,Hrycyna:2015eta,UrenaLopez:2011ur,Urena-Lopez:2015odd,Alho:2017opd,Hrycyna:2020jmw,Khyllep:2021yyp}. Yet, in that approach the dynamical system does not immediately close for an arbitrary potential, and one typically needs to resort to supplementary quantities or variables which complicates the situation.
Many authors prefer to work directly on the $(\phi, \dot \phi)$ plane \cite{Grain:2017dqa,Remmen:2013eja,Hergt:2018crm,Amendola:1990nn,Hrycyna:2008gk,Jarv:2008eb,Jarv:2010zc,Arefeva:2012sqa,Skugoreva:2014gka,Pozdeeva:2016cja,Kerachian:2019tar,Mishra:2019ymr,Tenkanen:2020cvw} which can be made compact using Poincar\'e transformation. However, this choice 
leads to visual difficulties for the following reason. The cosmological dynamical systems, especially in the more intricate case of nonminimal coupling, usually contains several singular asymptotic regimes where the scale factor and the scalar field evolve as some different powers of time. Clearly, if the field evolves as $\phi \sim t^{\alpha}$, where $t=0$ marks the cosmological singularity, the time derivative tends to infinity with greater rate $\dot \phi \sim t^{\alpha-1}$. Hence after the  Poincar\'e compactification of the phase space {\it any} such power-law trajectory converges to the {\it same} point in the asymptotic limiting circle. This property leads to difficulties in reading the phase diagrams when the dynamical system contains more than one such asymptotic regime. 
To overcome this problem and still keep intuitively clear variables we use in the present paper the set $(\phi, z)$, where $z=\tfrac{\dot \phi}{H}=\tfrac{d \phi}{d \ln a}$  \cite{Dutta:2020uha}. Very conveniently the derivative of the scalar field with respect to the number of expansion e-folds has the same dimension as $\phi$, the asymptotic power law regimes can be distinguished, and the dynamical system closes for any potential. For general purposes this choice is not entirely free of problems either. It can not be applied to any dynamics with bounces or recollapses since $z$ diverges at such points. Another disadvantage is the behavior of the scalar field near the origin during oscillations around the minimum of power law potentials $V\sim \phi^{2n}$. These oscillations are represented as usual circles on a $(\phi, \dot \phi )$ plane, but get extremely elongated in the $(\phi, z)$ variables due to the division by the small value of the Hubble parameter at the minimum of the potential. However, for the inflationary regime and the asymptotics of large $\phi$ these variables give us a clear picture of the dynamics.

Besides the general visual presentation, another goal of the current paper is to use the phase portraits to assess some quantitative characteristics of inflation.
The trajectory of inflationary Universe is subjected to some conditions in order to match the observable physical values. The amplitude of scalar and tensor perturbations, as well as the spectral tilts are almost completely determined by the parameters of the theory, but the number of e-folding $N$ during accelerated expansion is a characteristic of a particular trajectory. For physical reasons $N$ should be no less than approximately $50-60$. Studies of the initial conditions giving inflation with appropriate number of e-folds have been done since early days of inflation theory \cite{Belinsky:1985zd}. In these studies it was assumed that initially the scalar field has a Planckian density. This is quite reasonable for a minimally coupled scalar field since in this case the Planckian energy is a natural limit before quantum gravity should have an effect. However, when we turn to nonminimally coupled scalar fields, the Planck mass becomes dynamical and there is no reason to consider initial conditions using the contemporary value of it \cite{Gorbunov:2014ewa}. In Ref.\  \cite{Mishra:2019ymr} a zone of appropriate initial conditions was constructed at the $(\phi, \dot \phi)$ plane without fixing the initial energy of the scalar field. But this approach has its own disadvantages since different points in such zones could correspond to the same trajectory. Marking the zones of good initial conditions directly on the phase portraits helps to visualize the zone in question as a set of trajectories, not just points.

The analytical method to describe the slow motion of the scalar field was first introduced in the context of an isotropic bouncing model where it works equally well both during the contracting and expanding stages with an  exponential-like evolution of the scale factor \cite{1978SvAL....4...82S}.
In the context of inflation it is known as the slow-roll approximation \cite{Steinhardt:1984jj,Liddle:1994dx}. Concerning models with nonminimal coupling to curvature, there has been a simmering debate on how to properly define the slow roll conditions in the Jordan frame. On the one hand there is a proposal of generalized slow roll   \cite{Torres:1996fr,Faraoni:2000nt,Morris:2001ad,Chiba:2008ia,vandeBruck:2015gjd,Kuusk:2016rso,Akin:2020mcr}, while at the same time many researchers prefer to work in the Einstein frame from where the formulae can be translated back to the Jordan frame \cite{Chiba:2008ia,Kuusk:2016rso,Akin:2020mcr}.
As a byproduct of our investigations, we illuminate the issue. From the Jordan frame equations we derive a simple expression for a curve which happens to coincide with the Einstein frame slow roll condition translated into the Jordan frame. In the phase space this curve emerges from the (asymptotic) de Sitter fixed point tangentially to the inflationary eigendirection, and then closely approximates the ``attractor solution''. On the phase portraits it becomes apparent that this curve gives a better approximation to the inflationary trajectory than the generalized slow roll proposal.

While it is quite common to study inflation in nonminimal models by conformally transforming into the Einstein frame, the present work focuses upon the Jordan frame. We consider the Jordan frame to be physical in the sense that the physical lengths are determined in this frame. Although many Einstein frame results can be readily translated into the Jordan frame \cite{Jarv:2015kga,Jarv:2016sow}, there could be contexts where having the original Jordan frame description at hand would be beneficial. The physical effective Planck mass, which is relevant in the discussions of the initial conditions of inflation, is defined and its dynamics is clear in the Jordan frame, while in the Einstein frame its variability is hidden among the matter couplings and thus hard to track. Also, the count of the number of expansion e-folds as well as the end of inflation condition are slightly different in the Jordan and Einstein frames, and the precise determination of the end of inflation will affect the subleading predictions of the inflationary observables \cite{Karam:2017zno}. Besides, nonminimally coupled scalar field could be an effective description of a fundamental but more complicated theory, whereby transformation to the  Einstein frame is not obvious, see e.g.\  \cite{Koshelev:2020xby}. Furthermore, our approach can be generalized to other nonminimal theories which may not have a clear cut Einstein frame, e.g. the scalar field nonminimally coupled to the torsion \cite{Geng:2011aj,Hohmann:2018rwf} or nonmetricity \cite{Jarv:2018bgs} in the teleparallel setting.
Finally, let us note that the region of the phase space where the effective field theory for inflation is valid can arguably not reach infinity. However, since inflation is not a local phenomenon in the phase space that could be studied by focusing upon a vicinity of a certain point, but rather a global phenomenon happening along an extent of the leading phase trajectory, the information provided by the full phase portrait is still useful and illuminating.

The structure of the paper is as follows. In the next Sec.\ \ref{sec:Inflation in JF} we recall the key formulae to describe the cosmology of a nonminimally coupled scalar field in the Jordan frame and establish the slow roll conditions. In Sec.\ \ref{sec: dynamical system} we present the equations as a dynamical system and discuss the generic features of the phase space. As  particular illustrative examples we consider quadratic potential in Sec.\ \ref{sec: quadratic potential} and quartic potential in Sec.\ \ref{sec: quartic potential}, carrying out finite and infinite analysis, and presenting the results as the diagrams of Fig.\ \ref{quadratic_plots} and \ref{quartic_plots}, respectively. Then in Sec.\ \ref{sec: asymptotic regimes} we explain how the asymptotic regimes discussed before in the literature are in perfect correspondence with the asymptotic fixed points we found and mapped in the system. Finally Sec.\ \ref{sec: conclusions} is devoted to the conclusions. There is also a minor \ref{sec: appendix} for supplementary plots.

\section{Inflation in the Jordan frame}
\label{sec:Inflation in JF}
Let us write the Jordan frame action of a scalar field $\Phi$ nonminimally coupled to curvature $R$ as
\begin{equation}
\label{action}
S = \tfrac{1}{2}
\int d^4x\sqrt{-g}\left\lbrace F(\Phi) R-
\partial_\mu \Phi \, \partial^\mu \Phi - 2{V}(\Phi)\right\rbrace 
\,,
\end{equation}
Here $F(\Phi)$ is a nonminimal coupling function which plays the role of the dynamical reduced Planck mass squared, and sets the strength of the effective gravitational ``constant''.
In the following we assume attractive gravity with $F>0$, and that the potential is nowhere negative, $V(\Phi)\geq 0$. Hence the theory is free of ghosts, i.e. 
\begin{align}
\label{eq: no ghosts}
E(\Phi)=2F+3F_{,\Phi}^2 > 0 \,.
\end{align}
Here a comma denotes a derivative w.r.t.\ the scalar field, $F_{,\Phi}=\tfrac{d F(\Phi)}{d \Phi}$.
The condition \eqref{eq: no ghosts} can be observed e.g.\ in the Einstein frame where the dynamics of the metric and the scalar field are explicitly decoupled, as giving the correct sign of the scalar kinetic term in the action \cite{Boisseau:2000pr,EspositoFarese:2000ij,Jarv:2014hma}.

In a flat Friedmann-Lema\^{i}tre-Robertson-Walker background,
\begin{equation}
    ds^2=-dt^2+a^2(t) \, d\mathbf{x}^2 \,,
\end{equation}
the field equations arising from \eqref{action} and describing the evolution of the cosmological scale factor $a(t)$ in terms of the Hubble function $H=\tfrac{\dot{a}}{a}$ are
\begin{align}
\label{eq:Friedmann1}
3 F H^2 &= \frac{\dot{\Phi}^2}{2} - 3 F_{,\Phi} H \dot\Phi + V \,, \\
\label{eq:Friedmann2}
2 F \dot{H} &= -F_{,\Phi\Phi} \dot{\Phi}^2 + F_{,\Phi} H \dot{\Phi} - F_{,\Phi} \ddot{\Phi} - \dot{\Phi}^2 \,, \\
\label{eq:Scalar field equation}
\ddot{\Phi} + 3 H \dot{\Phi} &= 3 F_{,\Phi} \left(2 H^2 + \dot{H}\right) - V_{,\Phi} \,.
\end{align}
The rate of expansion can be conveniently expressed by the effective barotropic index 
\begin{align}
\weff &= -1 - \frac{2 \dot{H}}{3 H^2} 
\nonumber \\ &
= -1 + \frac{2 F_{,\Phi\Phi} \dot{\Phi}^2 - 2 F_{,\Phi} H \dot{\Phi} + 2 F_{,\Phi} \ddot{\Phi} + 2 \dot{\Phi}^2}{\dot{\Phi}^2 -6 F_{,\Phi} H \dot{\Phi} + 2 V} \,,
\label{eq: weff}
\end{align}
where the values $\weff<-\tfrac{1}{3}$ correspond to an accelerated expansion, while $\weff=-1$ is characteristic of de Sitter state where the scalar field abides at a fixed value at nonvanishing potential and behaves like a cosmological constant. In the minimally coupled case the barotropic index is bounded by $-1\leq \weff \leq 1$, but the expression \eqref{eq: weff} shows that in the nonminimally coupled case it is also possible to encounter superaccelerated ($\weff<-1$) as well as superstiff ($\weff>1$) rates of expansion. 

We may combine the equations \eqref{eq:Friedmann1}-\eqref{eq:Scalar field equation} to remove $\dot{H}$ in the scalar field equation, yielding
\begin{align}
\label{eq: ddotphi}
\ddot{\Phi} + 3H\dot{\Phi} + \frac{F_{,\Phi}\dot{\Phi}^2}{E}(1+3F_{,\Phi\Phi}) &= -\frac{2F}{E} \left(V_{,\Phi} - \frac{2F_{,\Phi}V}{F} \right) \,.
\end{align}
A lot about the dynamics can be understood by reading the RHS of \eqref{eq: ddotphi} as proportional to the gradient of the effective potential \cite{Chiba:2008ia,Skugoreva:2014gka,Jarv:2016sow},\footnote{Interestingly, a similar construction can be extended to models which add a nonminimal coupling to the Gauss-Bonnet invariant also \cite{Pozdeeva:2019agu,Pozdeeva:2021iwc}.}
\begin{align}
V_{\mathrm{eff}} = \frac{V}{F^2} \,.
\label{eq: Veff}
\end{align}
This quantity remains unchanged under conformal transformations and scalar fireld reparametrisations, and is thus an invariant characteristic of the scalar field dynamics \cite{Jarv:2014hma}. In particular, since $F$ and $E$ are both positive, the de Sitter fixed points where evolution of the scalar field stops, occur at the extrema of the effective potential, $V_{\mathrm{eff}}{}_{,\Phi}=0$.

To generate the nearly scale invariant spectrum of perturbations, the inflationary expansion must be nearly de Sitter like, which implies a slowly varying regime for the scalar field.
%
We may turn to the literature where one encounters the proposal of generalized slow roll conditions \cite{Torres:1996fr,Faraoni:2000nt,Morris:2001ad,Chiba:2008ia,vandeBruck:2015gjd,Kuusk:2016rso,Akin:2020mcr}. This approach assumes the usual conditions i) $\dot{\Phi}{}^2 \ll V$ and ii) $|\ddot{\Phi}| \ll H |\dot{\Phi}|$ of slow roll in minimally coupled theories, and complements those by further extended conditions iii) $|\ddot{F}| \ll H |\dot{F}|$ and iv) $H |\dot{F}| \ll H^2 F$. Applying the conditions i) and iv) on the Friedmann equation \eqref{eq:Friedmann1}, while in the Klein-Gordon equation \eqref{eq:Scalar field equation} assuming the condition ii) 
and completely neglecting the $\dot H$ term  
(which would follow from comparing Eqs.\ \eqref{eq:Friedmann1}, \eqref{eq:Friedmann2} under the conditions i), iii), iv) naturally)
results in 
\begin{subequations}
\label{eq: slow roll wrong}
\begin{align}
3FH^2 &\simeq V \,, \\
3H\dot{\Phi} &\simeq - V_{,\Phi} + \frac{2F_{,\Phi}V}{F} \,.
\end{align}
\end{subequations}

On the other hand it is known that translating the slow roll conditions from the Einstein to the Jordan frame results in a different set of equations \cite{Akin:2020mcr}:\footnote{In the specific case of induced gravity ($F\sim \Phi^2$) the conditions \eqref{eq: slow roll true} were derived in the Jordan frame in Ref.\ \cite{Kaiser:1993bq}.}
\begin{subequations}
\label{eq: slow roll true}
\begin{align}
\label{eq: slow roll true Friedmann}
3FH^2 &\simeq V \,, \\
\label{eq: slow roll true scalar}
3H\dot{\Phi} &\simeq -\frac{2F}{E} \left(V_{,\Phi} - \frac{2F_{,\Phi}V}{F} \right) \,.
\end{align}
\end{subequations}
We can get the equations \eqref{eq: slow roll true} directly from the Jordan frame equations if we do not completely omit the $\dot H$ term in the equation \eqref{eq:Scalar field equation}. In this case the structure of equations
of motion is quite different from the situation above, since now the system is not resolved with respect to highest derivative terms. 
To address the scalar field equation we should  use \eqref{eq: ddotphi} which can be rewritten as
\begin{align}
\label{eq: Scalar field equation rewritten}
    F_{,\Phi} \left( 3 \ddot{F} + 9 H \dot{F} + \dot{\Phi}{}^2 -4 V \right) &= -2F \left( \ddot{\Phi} + 3 H \dot{\Phi} + V_{,\phi} \right) \,.
\end{align}
Here at the LHS the condition iii) favors the second term over the first, while the condition i) gives importance to fourth term over the third. At the RHS the condition ii) suggests dropping the first term by taking it to be much smaller than the second. It is easy to check that combining the terms that remain at this point reproduces Eq.\ \eqref{eq: slow roll true scalar}. The impact of nonzero $\dot{H}$ is keeping the extra $\dot{F}\sim \dot{\Phi}$ term at LHS, since it could be of the same order as the $\dot{\Phi}$ term at RHS in the approximation. This can be seen as an effect of nonminimal coupling.

We should note that the additional factor $2 F/E$ in the second slow roll equation \eqref{eq: slow roll true scalar} depends only on $\Phi$ for a given $F$ but not on the details of time evolution of the cosmological system. It means that this factor does not
``feel'' how actually $\dot H$ is small with respect to $H^2$. All is needed that $\dot H$ is not equal to zero exactly, so that there is no continuous limit for  $\dot H/H^2 \to 0$. 

Both slow roll proposals \eqref{eq: slow roll true} and \eqref{eq: slow roll wrong} are obviously identical in the minimally coupled case. In the nonminimally coupled case they happen to coincide in the de Sitter fixed point limit, $V_{\mathrm{eff}}{}_{,\Phi}=0$, but start to differ for other values of $\Phi$.
In the subsequent sections we will compare the two proposals graphically on the phase portraits and see how the condition \eqref{eq: slow roll true} approximates quite well the leading inflationary trajectory, while the curve of condition \eqref{eq: slow roll wrong} can wander wildly off the mark and miss the actual inflationary dynamics, i.e.\ the ``attractor solution'' of the full nonlinear equations \eqref{eq:Friedmann1}-\eqref{eq:Scalar field equation}. As emphasised already for the minimally coupled case  \cite{Liddle:1994dx}, coming close to the true ``attractor solution'' is behind the success of the slow roll formalism.

\section{Dynamical system}
\label{sec: dynamical system}

So far we have not specified the function $F(\Phi)$. From now on we adopt the
following condition: $F(0)=M_{pl}^2$ where $M_{pl}$ is the Planck mass. This condition
ensures that we have the Einstein limit with $\Phi \to 0$. After that we are able 
to introduce 
 the following dimensionless variables for the dynamical system \cite{Dutta:2020uha}:
\begin{align}
\label{eq:dynamical variables}
\phi=\frac{\Phi}{M_{pl}} \,, \qquad z=\frac{\dot{\phi}}{H} \,.
\end{align}
It is useful to measure the evolution of the universe in dimensionless e-folds $N=\ln a$, using ${}'=\tfrac{d}{dN} = \tfrac{1}{H}\tfrac{d}{dt}$ as derivative. In particular, this implies $\phi'=z$. The number of expansion e-folds along a particular trajectory in the phase space can be also calculated easily as 
\begin{equation}
N = \int dN = \int H dt = \int \frac{H}{\dot{\phi}} d\phi = \int \frac{1}{z} d\phi \,.
\end{equation}
Here we assume expanding universe, $\dot{H}>0$, and note that $N>0$ for both increasing and decreasing $\phi$, since in the latter case also $z<0$. 

In these variables the set of cosmological equations \eqref{eq:Friedmann1}-\eqref{eq:Scalar field equation} is equivalent to a the dynamical system \cite{Dutta:2020uha}
\begin{align}
\phi' &= z \label{eq: dynsys phi}\\
z' &= -\frac{6F}{E} \left(F\frac{V_{,\phi}}{V} - 2F_{,\phi} \right) 
\nonumber \\ & \quad 
-\frac{3z}{E} \left( 3 F F_{,\phi} \frac{V_{,\phi}}{V} -3F_{,\phi}^2 +2F \right)
\nonumber \\ & \quad 
+ \frac{z^2}{E} \left(F \frac{V_{,\phi}}{V} - 3 F_{,\phi}^2 \frac{V_{,\phi}}{V} - 3 F_{,\phi\phi} F_{,\phi} - 7 F_{,\phi}\right) 
\nonumber \\ & \quad 
+ \frac{z^3}{2E} \left(F_{,\phi} \frac{V_{,\phi}}{V} + 2F_{,\phi\phi}  + 2\right) \,.
\label{eq: dynsys z}
\end{align}
This system would exhibit a singularity at the zeroes of the potential $V$, or become indeterminate if the potential has a minimum there as well (as even power law potentials do). Here it is helpful to assume that the potential is endowed with a tiny positive constant term that avoids the singularity of the system. On physical grounds we can consider this constant as a representation of late universe dark energy that is insignificant in comparison to the characteristic energy scale of early universe inflation. We will treat this constant as a regularizing parameter which can be made arbitrarily small.
Another notable feature of this dynamical system is that for the power law potentials without the constant term the coefficient in the potential (e.g. mass square of quadratic potential) will cancel out in \eqref{eq: dynsys z}.

From the Friedmann equation \eqref{eq:Friedmann1} we get a constraint
\begin{equation}
\label{eq: Friedmann constraint}
6F + F_{,\phi}z - z^2 \geq 0 
\end{equation}
if $V\geq 0$. This can be understood as putting a limit on the possible values of $z$ and bounding the physically allowed region of the phase space. In the minimally coupled case the bound is simply $-\sqrt{6} \leq z \leq \sqrt{6}$ independent of the value of $\phi$, while for nonminimal coupling the allowed domain of $z$ depends on the precise form of the coupling function $F$, and varies with $\phi$. Note that the constraint does not depend on the particular form of the potential, only on the assumption of nonnegativity of it. Another way to read \eqref{eq: Friedmann constraint} is to note that the constraint is exactly satisfied in the limit where the scalar field kinetic energy dominates over the potenital energy. Hence the boundary corresponds to the utmost regime of kinetic dominance.

The effective barotropic index in the phase space variables is
\begin{align}
\weff &= -1 + \frac{2F_{,\phi}}{E}\left(2 F_{,\phi} - \frac{F V_{,\phi}}{V} \right)  
+ \frac{z F_{,\phi}}{E} \left( \frac{2 F_{,\phi} V_{,\phi}}{V} + \frac{8}{3} \right) 
\nonumber \\ & \quad 
- \frac{z^2}{E} \left( \frac{2 F_{,\phi\phi}}{3} + \frac{F_{,\phi} V_{,\phi}}{3V}  - \frac{2}{3} \right) \,.
\label{eq: dynsys weff}
\end{align}
Again, in the minimal coupling case $\weff=-1+\tfrac{z^2}{3}$ and thus the rate of expansion can be directly read off from the respective phase portrait as a distance from the $z=0$ axis. The Friedmann constraint \eqref{eq: Friedmann constraint} ensures that at maximum $\weff=+1$, i.e.\ the stiff fluid regime as expected for kinetic dominance. In the nonminimal case, the phase portraits are less intituitive to understand as the effective barotropic index shows a complicated dependence on $\phi$ as well.

Demanding $\phi'=z'=0$ in the dynamical system \eqref{eq: dynsys phi}-\eqref{eq: dynsys z}, it is immediately possible to find the fixed points $(\phi=\phi_*, z=0)$, occuring at the values of $\phi_*$ such that
\begin{equation}
\frac{6F}{E} \left(2 F_{,\phi} - \frac{F V_{,\phi}}{V}\right)\Big|_{\phi_*} = 0 \,.
\label{eq: dynsys fixed point}
\end{equation}
Comparing with \eqref{eq: dynsys weff} tells that at the fixed points $\weff=-1$, i.e.\ the expansion is of the de Sitter type. 
The eigenvalues of the perturbed matrix at the fixed point are \cite{Dutta:2020uha}
\begin{equation}
\lambda_\pm =
-\frac{3}{2}\pm \frac{3}{2} \sqrt{1+\frac{8}{3E} \left(2F F_{,\phi\phi} + 2 F_{,\phi}^2 - F^2 \frac{V_{,\phi\phi}}{V} \right) } \,,
\end{equation}
hence these points can be attractors or saddles, depending on whether both eigenvalues have negative real parts or not. A closer inspection reveals that such fixed point is an attractor if it corresponds to a local minimum of the effective potential \eqref{eq: Veff}, and a saddle if it corresponds to a local maximum of the effective potential \cite{Dutta:2020uha}.

The slow roll approximation \eqref{eq: slow roll true}
can be translated into the dynamical variables as
\begin{equation}
\label{eq: slow roll true z}
z= \frac{2F}{E} \left( 2F_{,\phi} -\frac{F V_{,\phi}}{V} \right). 
\end{equation}
It draws a specific curve $z(\phi)$ in the phase space. This curve itself is not a phase trajectory (a solution of the full dynamical system), but approximates quite closely the leading solution which attracts all neighbouring trajectories. By comparing \eqref{eq: dynsys fixed point} and \eqref{eq: slow roll true z} it is straightforward to see that at $z=0$ the slow roll curve starts at a fixed point and then evolves away from it towards increasing $|z|$. As we observe later in the examples, the leading trajectory is none other but a heteroclinic orbit running from the saddle fixed point to the attractive fixed point. Inflationary expansion occurs at the initial part of this heteroclinic orbit, and it is the key for a successful description of inflation that the slow roll curve approximates well the leading solution. Later the middle part of the heteroclinic orbit corresponds to the scalar field oscillations responsible for reheating, while the final stage corresponds to settling down to the later eras involving dark energy. In the minimal coupling case substituting the slow roll curve \eqref{eq: slow roll true z} into the dynamical equation \eqref{eq: dynsys z} reveals that the slow roll curve follows $z'=0$. In the nonminimally coupled case a similar manipulation does not lead to an analogously simple result.

The alternative extended slow roll proposal \eqref{eq: slow roll wrong}
is also giving a phase space curve
\begin{equation}
\label{eq: slow roll wrong z}
z= 2F_{,\phi} -\frac{F V_{,\phi}}{V}
\end{equation}
which similarly starts from the fixed point at $z=0$. However, later the explicit plots show that the former curve \eqref{eq: slow roll true z} is a much better approximation to the leading trajectory. 

The behavior of the system at $\phi$ infinity can be revealed with the help of the Poincar\'e compactification
\begin{align}
\label{eq: Poincare variables}
\phi_p = \frac{\phi}{\sqrt{1+\phi^2+z^2}}\,, \qquad
z_p = \frac{z}{\sqrt{1+\phi^2+z^2}} \,.
\end{align}
The inverse relation between the compact variables $\phi_p, z_p$ and the original ones \eqref{eq:dynamical variables} is
\begin{align}
\phi=\frac{\phi_p}{\sqrt{1-\phi_p^2-z_p^2}} \,, \qquad z=\frac{z_p}{\sqrt{1-\phi_p^2-z_p^2}} \,.
\end{align}
By construction the compact variables are bounded, 
$\phi_p^2+z_p^2\leq 1$, and map an infinite phase plane onto a disc with unit radius. We can express the dynamical system along with the fixed points, slow roll curves, etc.\ in terms these new variables, to get a compact global picture of the full phase space including the asymptotic regions.

\section{Quadratic potential}
\label{sec: quadratic potential}

To draw the phase portraits we need to fix the model functions. First let us take quadratic nonminimal coupling and quadratic potential,
\begin{equation}
\label{eq: quadratic model}
F = 1 + \xi \phi^2 \,, \qquad V= \frac{m^2}{2} \phi^2 + \Lambda \,.
\end{equation}
As explained above, we consider $\Lambda$ as a small regularizing parameter, to avoid the system \eqref{eq: dynsys phi}-\eqref{eq: dynsys z} blow up at the origin. We compute all quantities with $\Lambda>0$ and then apply the limit $\Lambda \rightarrow 0$ to present the formulae and draw the plots on Fig.\ \ref{quadratic_plots} for four qualitatively representative cases of $\xi=0, \, 0.005, \, 1, \, 10000$. 

\subsection{Finite analysis}

\begin{figure*}
	\centering
	\subfigure[]{
		\includegraphics[width=5.5cm]{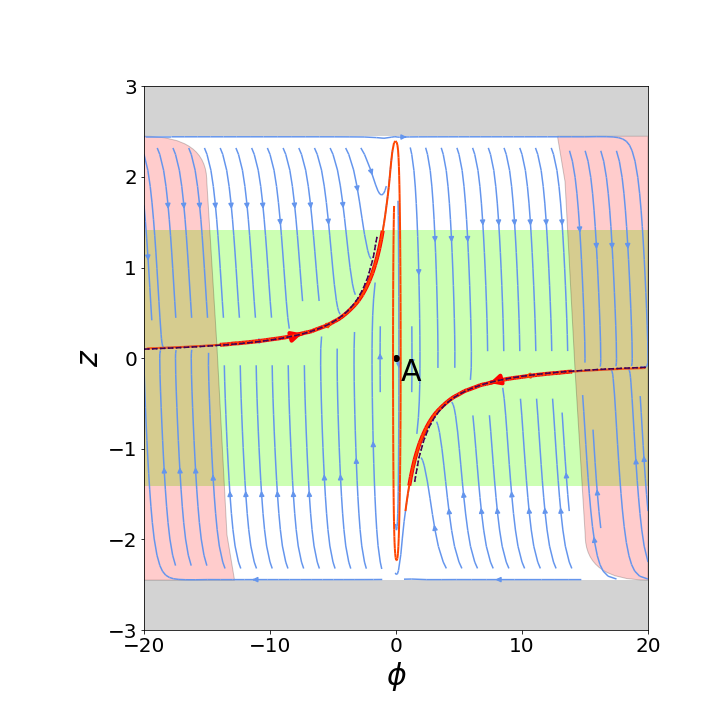} \label{fig: quadratic_xi_0_finite}}
	\subfigure[]{
		\includegraphics[width=5.5cm]{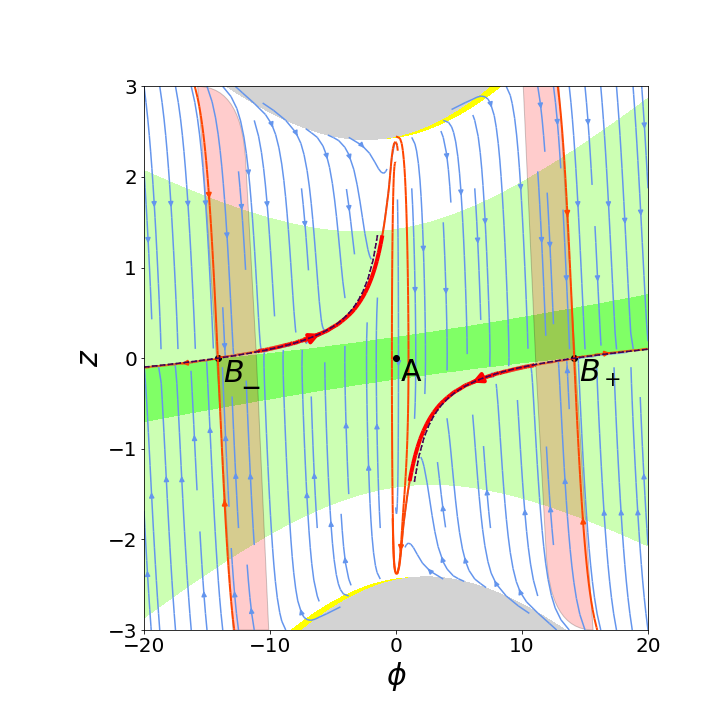} \label{fig: quadratic_xi_0.005_finite}}
	\subfigure[]{
		\includegraphics[width=5.5cm]{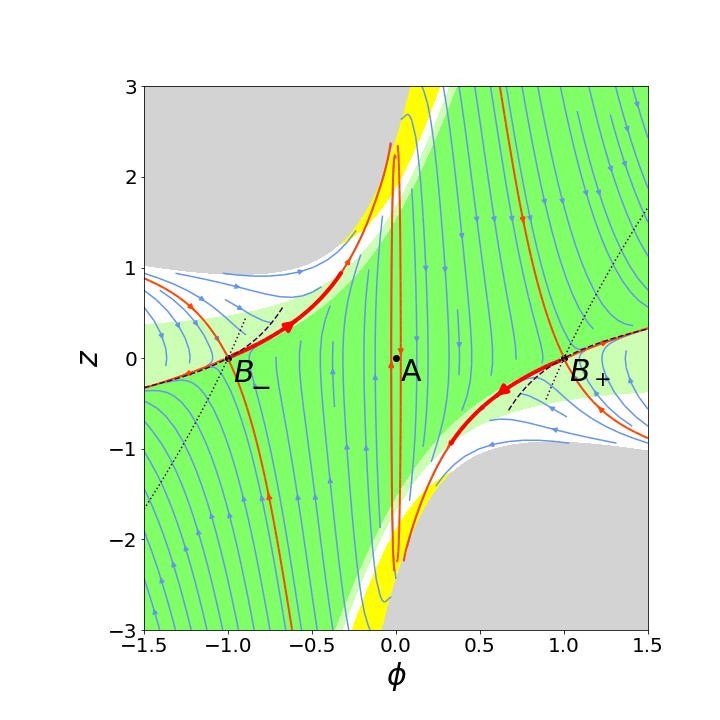} \label{fig: quadratic_xi_1_finite}}
\\
	\subfigure[]{
		\includegraphics[width=5.5cm]{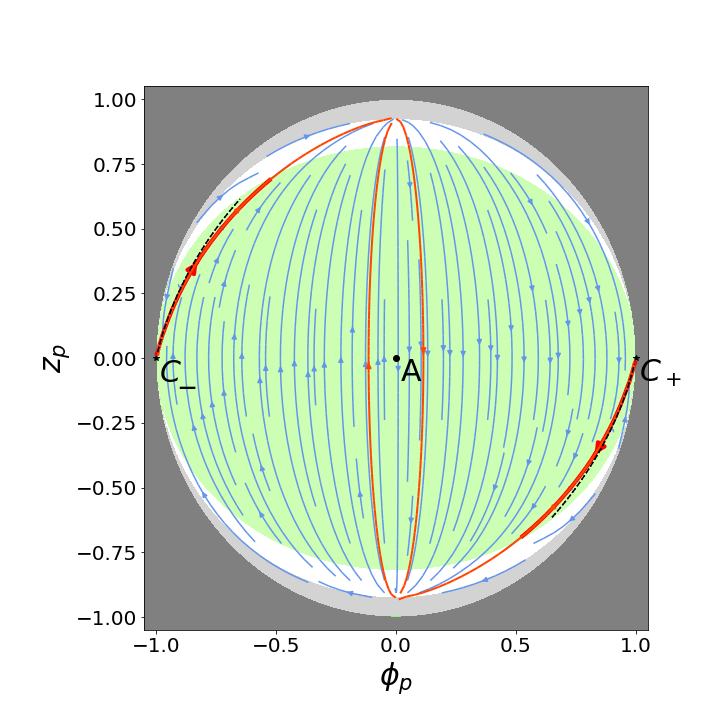} \label{fig: quadratic_xi_0_infinite_full} }
    \subfigure[]{
		\includegraphics[width=5.5cm]{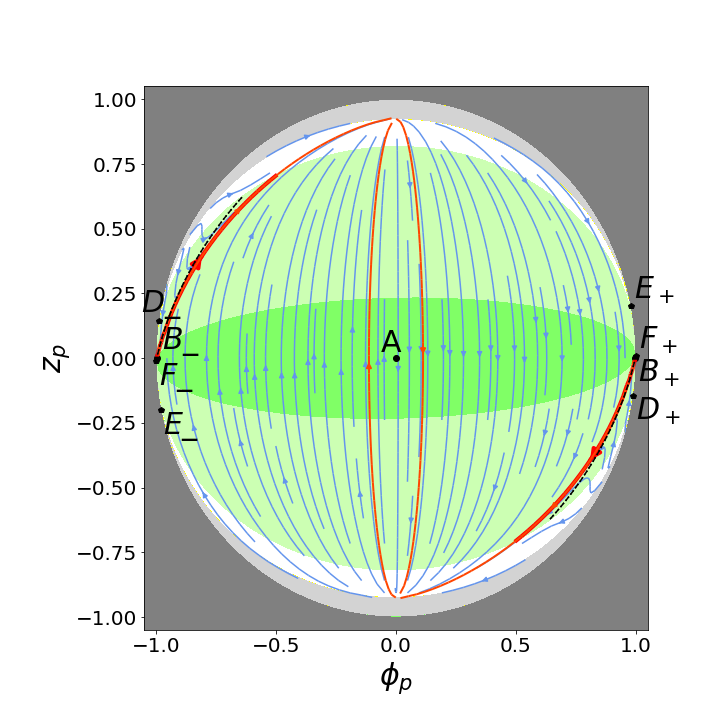} \label{fig: quadratic_xi_0.005_infinite_full} }
	\subfigure[]{
		\includegraphics[width=5.5cm]{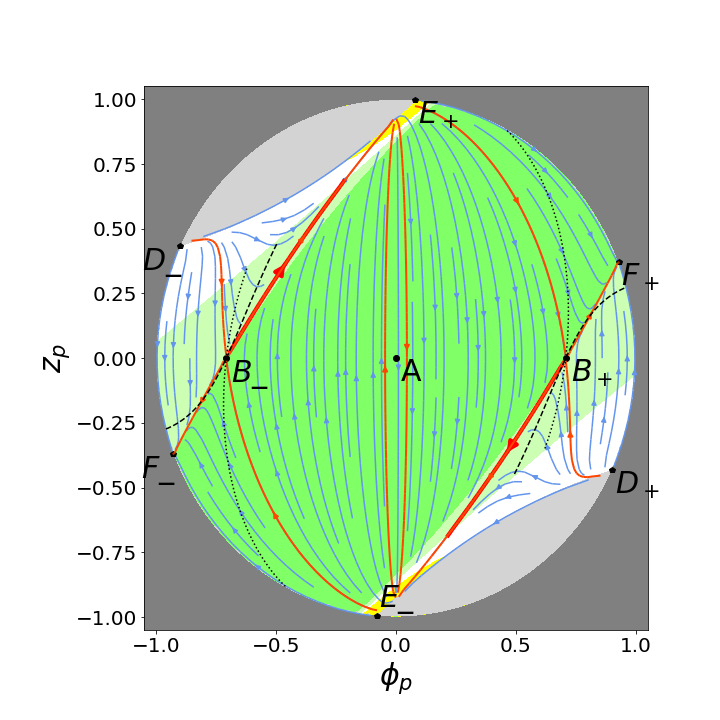} \label{fig: quadratic_xi_1_infinite_full}}
\\
	\subfigure[]{
		\includegraphics[width=5.5cm]{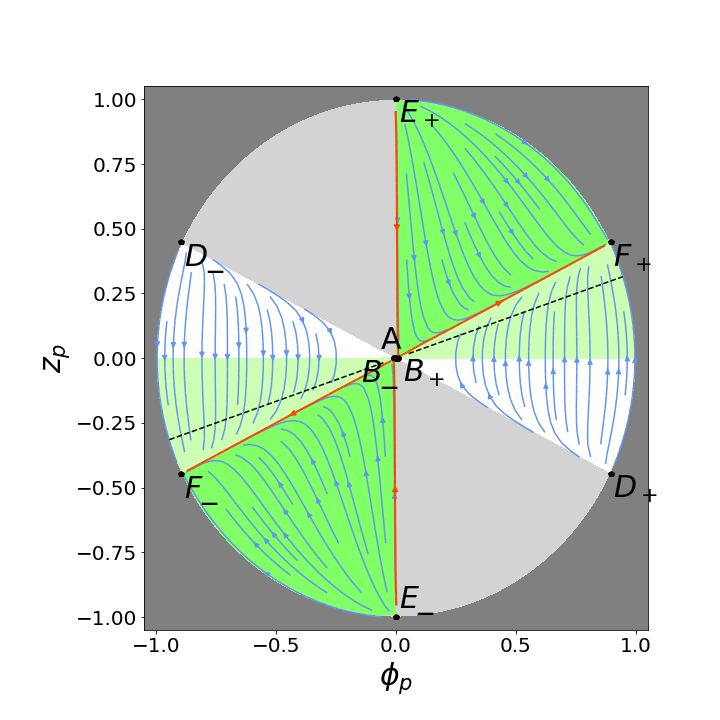} 
		\label{fig: quadratic_xi_10000_infinite_full}}
	\subfigure[]{
		\includegraphics[width=5.5cm]{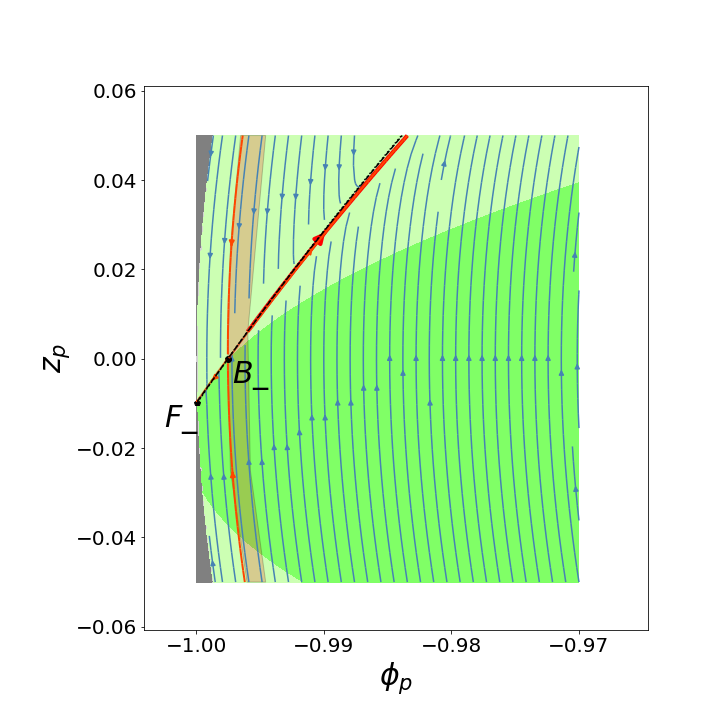} 
		\label{fig: quadratic_xi_0.005_infinite_corner}}
	\subfigure[]{
		\includegraphics[width=5.5cm]{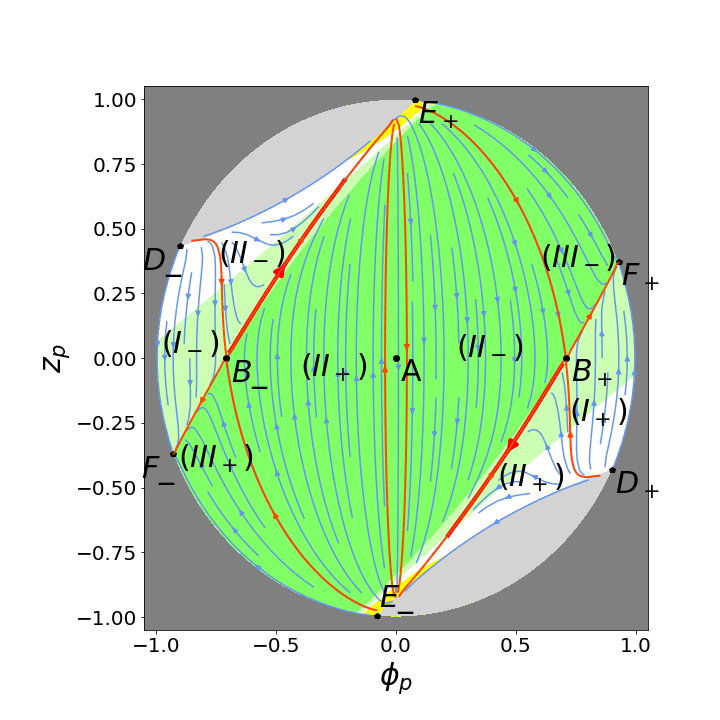} 
		\label{fig: quadratic_xi_1_infinite_full_regions} }
	\caption{Cosmological phase portraits of the quadratic model \eqref{eq: quadratic model} with $\Lambda=0$, and $\xi=0$ (panels a, d), $\xi=0.005$ (panels b, e, h), $\xi=1$ (panels c, f, i), $\xi=10000$ (panel g). Green background stands for superaccelerated, light green accelerated, white decelelerated, and yellow superstiff expansion, while grey covers the unphysical region. Orange trajectories are heteroclinic orbits between the fixed points, wider red part of the master trajectory highlights the 50 last e-folds of inflation, while red semi-transparent cover shows the basin of initial conditions leading to 50 e-folds. The dashed curve marks the path of slow roll approximation \eqref{eq: slow roll true z phi2}, while the dotted curve runs along the other slow roll approximation \eqref{eq: slow roll curve wrong phi2} suggested in the literature.}
	\label{quadratic_plots}
\end{figure*}

The dynamical system \eqref{eq: dynsys phi}, \eqref{eq: dynsys z} for the quadratic model \eqref{eq: quadratic model} reads
\begin{align}
\label{eq: dynsys phi phi2}
\phi' &= z \\
z' &= \frac{ (z^2 - 6) (\phi z + 2)}{2 \phi (1 + \xi \phi^2 + 6\xi^2 \phi^2 )} 
\nonumber \\ & \quad 
- \frac{ \xi (3\phi^2 z + 6\phi z^2 - 2 z^3 + 18 z) }{ (1 + \xi \phi^2 + 6\xi^2 \phi^2 )} - \frac{6 \xi^2 \phi (\phi^2 - 3z^2)  }{ (1 + \xi \phi^2 + 6\xi^2 \phi^2 )} \,.
\label{eq: dynsys z phi2}
\end{align}
As mentioned before, it is the nature of the system that the mass parameter $m$ cancels out when the constant term $\Lambda$ is neglected. The only remaining parameter that characterizes the system is the nonminimal coupling $\xi$. Also note that, the system is invariant under the diagonal reflection $\phi\rightarrow -\phi$, $z\rightarrow -z$, which follows from the symmetry of the potential and the nonminimal coupling function. The qualitative flow of generic solutions of the system is shown by the blue trajectories on the top panels of Fig.\ \ref{quadratic_plots}.

For the quadratic model \eqref{eq: quadratic model}, the barotropic index \eqref{eq: dynsys weff} is
\begin{align}
\weff &= -1 + \frac{z^2 + 4 \xi(z^2-2 \phi z -3) +12 \xi^2 \phi (\phi-2z)}{3 (1+ \xi \phi^2 + 6 \xi^2 \phi^2)} \,. 
\label{eq: dynsys weff phi2}
\end{align}
On Fig.\ \ref{quadratic_plots} the phase space area corresponding to accelerated expansion ($-1 \leq \weff < -\tfrac{1}{3}$) is painted by light green on the plots, while the zone of decelerated expansion is indicated by white background. In the case of nonminimal coupling we also encounter superaccelerated expansion ($ \weff < -1$) marked by dark green, and superstiff expansion ($ \weff > +1$) marked by yellow.

The physical phase space is bounded \eqref{eq: Friedmann constraint} by
\begin{align}
\label{eq: dynsys boundary quadratic}
z^2 - 6\xi \phi (\phi + 2z) < 6
\end{align}
One can check explicitly that the solutions do not cross this boundary, e.g.\ by computing the scalar product of the flow vector $(\phi',z')$ and a vector normal to the boundary, and seeing the product is zero. This also means there is a solution trajectory running along the boundary that blocks other solutions to reach the boundary everywhere in the phase space, except for the asymptotics.

In the minimally coupled case the layout of the phase space is very simple (see Fig.\  \ref{fig: quadratic_xi_0_finite}), as the lines of constant $\weff$ are horizontal, spanning from $\weff=-1$ at $z=0$ to the stiff expansion limit $\weff=+1$ at the boundary of the physical phase space at $z_b^\pm=\pm\sqrt{6}$.
 Nonminimal coupling distorts this straight picture, the boundary does not correspond to a fixed $\weff$, and superstiff behaviour is also possible near the boundary of the phase space (Figs.\  \ref{fig: quadratic_xi_0.005_finite}, \ref{fig: quadratic_xi_1_finite}). In particular, tracing the boundary
 \begin{align}
 \label{eq: quadratic z_b}
 z_{b}^\pm (\phi) &= 6\xi \phi \pm \sqrt{6+6\xi\phi^2 + 36\xi^2\phi^2}
 \end{align}
 where ``$+$'' corresponds to the upper boundary (maximal value of $z$) and ``$-$'' to the lower boundary (minimal value of $z$), 
 to the asymptotics $\phi\rightarrow \pm \infty$ we get the value of the barotropic index as
 \begin{align}
 \label{eq: quadratic w_eff_b}
\weff{}_{,b}^\pm &= \frac{3\xi+42\xi^2+144\xi^3\pm 4\sqrt{6}(\xi+6\xi^2)^{\frac{3}{2}}}{3\xi (1+6\xi)}
\end{align}
Here $\weff{}_{,b}^-$ corresponds to ``upper left'' $(-\infty,z_b^+)$ and ``lower right'' $(\infty,z_b^-)$ boundary asymptotic, and drops from the value $+1$ of minimal coupling ($\xi=0$) to $+\tfrac{1}{3}$ for infinite coupling ($\xi\rightarrow \infty$). Analogously, $\weff{}_{,b}^+$ corresponds to ``upper right'' $(\infty,z_b^+)$ and ``lower left'' $(-\infty,z_b^-)$ boundary asymptotic, it increases from the value $+1$ of minimal coupling to infinite value for inifinite coupling.  
 
There are three regular fixed points in general. First at the origin is 
\begin{align}
A &: (0, 0) 
\end{align}
which is always a stable focus. This point occurs at the local minimum of the effective potential \eqref{eq: Veff}, and corresponds to late time universe. If we maintain nonzero $\Lambda$, it exhibits regular de Sitter expansion, while for vanishing $\Lambda$ we get Minkowski. The energy scale of a cosmological constant or dark energy is negligible in comparison to the scale of inflation, and we are not concerned to follow the scalar field oscillations around the minimum into the dark energy era. Thus for our purposes it made complete sense to take $\Lambda\rightarrow 0$ for simplicity.

At both sides the point $A$ is accompanied by the other fixed points
\begin{align}
B_\pm &: (\pm \tfrac{1}{\sqrt{\xi}}, 0)
\end{align}
which are saddles, and correspond to a local maximum of the effective potential \eqref{eq: Veff}. In the limit where $\xi$ vanishes the effective potential $V_{\mathrm{eff}}$ reduces to $V$, these points are pushed to the asymptotics of $\phi$ and become nonhyperbolic. For nonminimal coupling the points $B_\pm$ correspond to regular de Sitter expansion of the universe, and on the plots \ref{fig: quadratic_xi_0.005_finite}, \ref{fig: quadratic_xi_1_finite} we can observe how they sit exactly at the border between accelerating (light green) and superaccelerating (green) regions of the phase space. In the limit of minimal coupling these points should be better called asymptotic de Sitter, to be studied more closely later below.

As typical for a saddle, the repulsive eigendirections of $B_\pm$ launch a pair of opposite outgoing trajectories (solutions) from the fixed point, which act attractively for the neighbouring trajectories. In fact, these master trajectories that exit $B_\pm$ towards decreasing $|\phi|$ are actually heteroclinic orbits as they end up at the other fixed point $A$. The second pair of master trajectories that exit $B_\pm$ towards increasing $|\phi|$ and run to infinity, turn out to be heteroclinic orbits as well, since there is another fixed point in the asymptotics to receive them, to be revealed by the global portraits below. Furthermore, the trajectories which enter $B_\pm$ along the saddle attractive eigendirections are also heteroclinic orbits, originating from yet another set of asymptotic fixed points. On the plots the heteroclinic orbits are indicated by orange trajectories, numbering four for both saddle points $B_\pm$.

Among these heteroclinic orbits, the ones which run from $B_\pm$ to $A$ are central for inflation, as they offer a phase space description of the phenomenon of an ``inflationary attractor'', i.e.\ why for a large class of initial conditions the dynamics converges close to a particular master solution that leads lots of trajectories with similar behaviour in accelerated expansion, and finally into an exit to late universe. Thus we see that a nonhyperbolic fixed point and the corresponding center submanifold \cite{Alho:2014fha,Alho:2015cza,Alho:2017opd}, are not the only possibility to realize inflation, the same can be achieved by a saddle point and unstable submanifold. 
The other pair of trajectories leading out of the $B_\pm$ fixed points are also attracting neighbouring trajectories to flow along with them, but as they run into infinity and there is no end to accelerated expansion, they cannot offer a viable option for inflation. 

The path along these inflationary master trajectories that corresponds to the last 50 e-folds of accelerated expansion before the end of inflation (reaching $\weff=-\tfrac{1}{3}$ from below) is highlighted by a wider red ribbon on the plots of Fig.\ \ref{quadratic_plots}. In the minimal and very small $\xi$ couplings the inflationary master trajectory can traverse some distance from the fixed point before the last 50 e-folds start, while as $\xi$ gets larger the points $B_\pm$ move ever closer to the origin and the last 50 e-folds tend to begin very close to the fixed point. 

The basin of the initial conditions which leads to at least 50 e-folds of expansion is covered by semi-transparent red hue on the plots \ref{fig: quadratic_xi_0_finite}, \ref{fig: quadratic_xi_0.005_finite}. For larger $\xi$ on the plot \ref{fig: quadratic_xi_1_finite} this basin is too narrow to be depicted, it is rather indistinguishable from the heteroclinic orbits which run into $B_\pm$. Although generic trajectories can experience roughly 1 e-fold of accelerated expansion before reaching the vicinity of the master trajectory, the most significant amount of inflation clearly occurs only when they closely follow the master solution. This happens because for the most part along the master solution $\phi$ evolves very slowly and the conditions are almost de Sitter, while off from the ``attractor solution'' the scalar field experiences quick evolution, sometimes called a ``fast roll.''

It is easy to check that the slow roll curve \eqref{eq: slow roll true z}, marked by a dashed black line on the plots, and given by
\begin{align}
\label{eq: slow roll true z phi2}
z &= - \frac{2 (1- \xi^2 \phi^4)}{\phi (1 + \xi \phi^2 +  6\xi^2 \phi^2)} \,
\end{align}
runs through the fixed points $B_\pm$. In fact, we may consider the $B_\pm$ as the starting points of the slow roll curves, since the tangent of the slow roll curves at these points matches the repulsive eigendirections of the saddles. Thus the slow roll curve \eqref{eq: slow roll true z phi2} approximates rather well the inflationary master trajectory, which explains the success of the slow roll conditons in the description of inflation. One should still be reminded that the slow roll curve is an approximation and not itself a solution of the dynamical equations \eqref{eq: dynsys phi phi2}-\eqref{eq: dynsys z phi2}. The slow roll curve slightly overestimates the actual evolution of the scalar field, and prescribes larger $|z|$ than the actual leading trajectory. As a result the slow roll curve exits the zone of accelerated expansion at higher values of $|\phi|$ than the ``attractor solution.'' This difference becomes more pronounced at larger nonminimal coupling $\xi$, as can be observed on the top panels of Fig.\ \ref{quadratic_plots}. (For minimal coupling the same effect can be noted on the plots in different variables in e.g.\ Refs.\ \cite{UrenaLopez:2007vz,Grain:2017dqa}.)

The alternative putative slow roll curve proposed in the literature \eqref{eq: slow roll wrong z} is given by
\begin{align}
\label{eq: slow roll curve wrong phi2}
z &=-\frac{2(1-\xi\phi^2)}{\phi} \,
\end{align}
and marked by a dotted line on the plots. It does also run through the fixed points $B_\pm$, but its tangent is not along the saddle repulsive eigendirection of $B_\pm$, and it gives a much worse, basically wrong approximation of the actual inflationary master solution, as is clearly seen on Fig.\ \ref{fig: quadratic_xi_1_finite}. As remarked before, in the limit of minimal coupling both proposals coincide and thus are hard to distinguish on Figs.\ \ref{fig: quadratic_xi_0_finite} and \ref{fig: quadratic_xi_0.005_finite}.

In the minimal coupling case the slight ($z \nleftrightarrow -z$) asymmetry in the initial conditions that guarantee at least 50 e-folds of accelerated expansion on Fig.\ \ref{fig: quadratic_xi_0_finite} has a simple physical explanation. The solutions at negative $\phi$ and positive $z$ run down the potential and if the initial ``speed'' $z$ is too high, they may slow down too late to experience sufficient slow roll regime. On the other hand the solutions at negative $\phi$ and negative $z$ initially climb up the potential, turn around at $z=0$ and start rolling down, thus having more chance to follow the slow roll regime. The logic of the phase space even asserts, that there must be a class of extreme solutions which come from the positive side of $\phi$ with negative $z$ in the kinetically dominated manner near the physical phase space boundary (too close to be marked explicitly), reach sufficiently high up in the potential, turn around, and consequently also experience enough e-folds of accelerated expansion. To be clear, though, these extreme kinetic solutions would have initial energy densities well beyond the Planckian limit, and are of questionable physical relevance. Thus the picture is in a qualitative agreement with the numerical investigations of Ref.\ \cite{Mishra:2019ymr}. 

In the case of nonminimal coupling slow roll occurs when the scalar field rolls down from the local maximum of the potential at the fixed point $B_\pm$ towards the central fixed point $A$. For instance, taking the point $B_-$ on Fig.\ \ref{fig: quadratic_xi_0.005_finite}, only those solutions can experience enough inflation which either cross over the potential maximum from the left at sufficiently low ``speed'' z, or which climb up towards the potential maximum from the right but do not shoot over the top. Both options have a rather limited range and thus the band of good initial conditions narrows down compared to the minimally coupled case. The situation becomes even more precarious at high nonminimal coupling $\xi$, because the fixed points are pushed closer to the center and the path of the possible slow roll gets shorter. Only very finely tuned initial conditions whereby the field evolution is really slow near the maximum of the potential, grant the privilege to experience at least 50 e-folds of accelerated expansion. Thus for quadratic potentials invoking nonminimal coupling severely limits the range of initial conditions conductive for proper inflation. 
Let us remark, that in the nonminimal coupling case there are also extreme kinetic solutions near the boundary of the physical phase space which cross over $\phi=0$ (too narrow to be shown explicitly). Among these there are some lucky trajectories which slow down precisely enough to reach the vicinity of $B_\pm$ and subsequently partake in the inflationary regime. The immediate assessment of the physicality of the initial conditions requires now extra caution, however, since the effective Planck mass is dynamical in the nonminimal coupling case. Due to the resulting ambiguity in delimiting the region of Planckian energy density, we have instead opted to delineate all trajectories which give sufficient inflation.

\subsection{Infinite analysis}

To unravel the asymptotic fixed points it is helpful to rewrite the system \eqref{eq: dynsys phi phi2}-\eqref{eq: dynsys z phi2} in terms of the compact variables \eqref{eq: Poincare variables}, yielding
\begin{align}
\phi_p' &= \frac{z_p (14 q^2 + 6 q \phi_p z_p - \phi_p z_p^3)}{2 (q + \xi \phi_p^2 + 6 \xi^2 \phi_p^2)} 
\nonumber \\ & \quad 
+ \frac{\xi \phi_p z_p (q \phi_p + 18 q z_p + 3 \phi_p^2 z_p + 7 \phi_p z_p^2 - 2 z_p^3)}{q + \xi \phi_p^2 + 6 \xi^2 \phi_p^2} 
\nonumber \\ & \quad 
+ \frac{6 \xi^2 \phi_p^2 z_p (q - \phi_p^2 + 4 z_p^2) }{q + \xi \phi_p^2 + 6 \xi^2 \phi_p^2} \\
z_p' &= -\frac{6 q^2 (q + \phi_p^2)}{\phi_p (q + \xi \phi_p^2 + 6 \xi^2 \phi_p^2 )} 
\nonumber \\ & \quad 
-\frac{ z_p ( 6 q^2 \phi_p - 2 q^2 z_p + 6 q \phi_p^3 - q \phi_p z_p^2 - \phi_p^3 z_p^2}{2 \phi_p (q + \xi \phi_p^2 + 6 \xi^2 \phi_p^2 )} 
\nonumber \\ & \quad 
- \frac{\xi z_p (18 q^2 + 21 q \phi_p^2 + 3 \phi_p^4)}{q+ \xi \phi_p^2  + 6  \xi^2 \phi_p^2 } 
\nonumber \\ & \quad 
- \frac{\xi z_p^2 (6 q \phi_p -2 q \phi_p z_p + 7 \phi_p^3 - 2 \phi_p^2 z_p)}{q+ \xi \phi_p^2  + 6  \xi^2 \phi_p^2 } 
\nonumber \\ & \quad 
+ \frac{6 \xi^2 \phi_p (q \phi_p^2 -3 q z_p^2 + \phi_p^4 - 4 \phi_p^2 z_p^2)}{q + \xi \phi_p^2 + 6  \xi^2 \phi_p^2 } \,.
\end{align}
Here we have introduced a shorthand expression $q=1-\phi_p^2-z_p^2$. In the compact variables the global dynamics fits into a unit circle, $\phi_p^2+z_p^2 \leq 1$, and is depicted on the lower panels of Fig.\ \ref{quadratic_plots}. Note that after the transformation the system still exhibits the diagonal symmetry $\phi_p\rightarrow -\phi_p$, $z_p \rightarrow -z_p$.

The barotropic index is expressed by
\begin{align}
\weff &= -1 \nonumber \\
&+ \frac{z_p^2+4\xi(3\phi_p^2 - 2 \phi_p z_p + 4 z_p^2 -3) + 12 \xi^2 \phi_p (\phi_p -2 z_p)}{3(q + \xi \phi_p^2 + 6 \xi^2 \phi_p^2)} \,,
\end{align}
while the color coding of different expansion regimes on the diagram is the same as before. The physical phase space is bounded by 
\begin{align}
\label{eq: unphysical pz phi2}
\frac{z_p^2- 6\xi \phi_p (\phi_p+2 z_p)}{q} < 6 \,.
\end{align}
In the minimal coupling case the physical phase space includes only the points $(\pm 1, 0)$ in the asymptotics, i.e.\ the asymptotic ``left'' edge of the phase space from $(-\infty,-\sqrt{6})$ to  $(-\infty,\sqrt{6})$ in the original noncompact variables is mapped to a single point $(-1, 0)$ in the compact variables, and analogously the ``right'' edge as well. In the nonminimal coupling case, the asymptotic edges do not get contracted into a point in the compact variables, i.e.\ the asymptotic ``left'' edge of the noncompact phase space from $(-\infty,z_b^-)$ to $(-\infty,z_b^+)$ is mapped to a stretch of the asymptotic circle. 

In the coordinates $(\phi_p,z_p)$ the finite fixed points are located at
\begin{align}
A &: (0, 0)
\end{align}
and
\begin{align}
B_\pm &: (\pm \frac{1}{\sqrt{1+\xi}}, 0) \,,
\end{align}
maintaining their characteristics. In the limit of minimal coupling the latter points are pushed to the asymptotics, and become
\begin{align}
C_\pm &: (\pm 1, 0) \,.
\end{align}
As the real part of the eigenvalues vanishes in the $\xi \rightarrow 0$ limit these points turn from saddles to nonhyperbolic. The effective barotropic index at $C_\pm$ is still equal to $-1$, thus featuring de Sitter like expansion, much alike the point described by Ref.\ \cite{Alho:2014fha} in other variables. But the scalar field approaches infinity meaning the effective gravitational ``constant'' vanishes while the potential is infinite. In Sec.\ \ref{sec: asymptotic regimes} we are able to identify the corresponding dynamics which is distinct from the regular de Sitter and warrants a name asymptotic de Sitter.

The slow roll curve \eqref{eq: slow roll true z phi2} is given in the implicit form as
\begin{align}
\label{eq: slow roll true pz phi2}
\frac{2q^2 +q\phi_p z_p + \xi \phi_p^3 z_p - 2\xi^2 \phi_p^3 (\phi_p - 6 z_p)}{\sqrt{q}\phi_p (q + \xi \phi_p^2 + 6 \xi^2 \phi_p^2)} = 0 \,,
\end{align}
while the alternative and less successful slow roll proposal  \eqref{eq: slow roll curve wrong phi2} translates to 
\begin{align}
\frac{2q +\phi_p z_p -2 \xi \phi_p^2}{\sqrt{q}\phi_p} = 0 \,.
\end{align}
The slow roll curves still start at $B_\pm$, or $C_\pm$, but only the tangent of the first curve \eqref{eq: slow roll true pz phi2} points in the repulsive eigendirections, and gives an acceptable approximation to the actual inflationary master solution. Similarly to the finite case, the global phase portraits show the path corresponding to the last 50 e-folds of inflationary expansion in broad highlighed red on the master trajectory. The basin of initial conditions which allow such expansion is rather narrow in the compact coordinates and is not shown on the global diagrams. A patch of this basin in the $\xi=0.005$ case is depicted on Fig.\ \ref{fig: quadratic_xi_0.005_infinite_corner} to illustrate the situation.

In the asymptotic limit $q\rightarrow 0$ the system reduces to
\begin{align}
\phi_p' &= - \frac{z_p (z_p- 2\xi (\phi_p-2z_p))(z_p^2- 6\xi \phi_p( \phi_p +2 z_p))}{ 2 \xi \phi_p (1+6\xi)} \\
z_p' &= \frac{(z_p- 2\xi (\phi_p-2z_p))(z_p^2- 6\xi \phi_p( \phi_p +2 z_p))}{ 2 \xi (1+6\xi)} \,,
\end{align}
where $z_p^2=1-\phi_p^2$. Looking for the asymptotic fixed points reveals that in the case of nonminimal coupling there are fixed points where the boundary of the physical phase space touches the circle of infinity:
\begin{align}
D_\pm &: (\pm\Omega_+,\mp \sqrt{1-\Omega^2_+}) \,, \\
E_\pm &: (\pm\Omega_-,\pm \sqrt{1-\Omega^2_-}) 
\end{align}
with
\begin{align}
\label{eq: Omega_pm}
\Omega_\pm &= \frac{\sqrt{1+6\xi+72\xi^2 \pm12\xi \sqrt{6\xi+36\xi^2}}}{\sqrt{1+12\xi+180\xi^2}}
\end{align}
The points $D_\pm$ correspond to the ``upper left'' $(-\infty,z_b^+)$ and ``lower right'' $(\infty,z_b^-)$ corners of the physical phase space in finite variables, and have the barotropic index $\weff{}_{,b}^-$ given in Eq.\ \eqref{eq: quadratic w_eff_b}. These points turn out to be unstable nodes, i.e.\ global sources for the bulk of the trajectories in the phase space. Here one repulsive eigendirection points tangentially along the boundary of the unphysical region \eqref{eq: unphysical pz phi2}, and the other repulsive eigendirection lies along the circle of infinity. Between these two directions there are heteroclinic orbits running from the points $D_\pm$ to the points $B_\pm$, as alluded before. One of these heteroclinic orbits goes ``directly'' from $D_\pm$ to $B_\pm$, while the other one passes through $\phi=0$ and ends up at $B_\mp$. The points $E_\pm$ correspond to the ``lower left'' $(-\infty,z_b^-)$ and ``upper right'' $(\infty,z_b^+)$ corners of the physical phase space in finite variables, and have the barotropic index $\weff{}_{,b}^+$ given in Eq.\ \eqref{eq: quadratic w_eff_b}. These points turn out to be saddle points with the attractive eigendirection pointing along the boundary of the unphysical region \eqref{eq: unphysical pz phi2} and repulsive eigendirection lying transverse to the circle of infinity. There are heteroclinic orbits running from the points $D_\pm$ to the points $E_\mp$ along the boundary of the unphysical region (not drawn explicitly on the plots). For nonminimal coupling there is yet another pair of asymptotic fixed points at
\begin{align}
F_\pm &: (\pm\Omega_F,\mp \sqrt{1-\Omega^2_F}) \,,
\end{align}
where
\begin{align}
\label{eq: Omega_F}
\Omega_F &= \frac{1+4\xi}{\sqrt{1+8\xi+20\xi^2}} \,.
\end{align}
The points $F_\pm$ are stable nodes and have barotropic index $\weff=-1$. One of the attractive eigendirections of $F_\pm$ is tangential to the circle of infinity, whereby it receives heteroclinic orbits coming from $D_\pm$ and $E_\pm$ (not drawn explicitly), while the other attractive eigendirection is oriented towards the interior of the phase diagram, whereby it receives the heteroclinic orbit from the point $B_\pm$. The physical regimes corresponding to these asymptotic fixed points are identified in Sec.\ \ref{sec: asymptotic regimes}.

To recap, for the nonminimal coupling the global picture follows a rather clear structure. The heteroclinic orbits connecting the fixed points divide the phase space into sectors, see Fig.\ \ref{fig: quadratic_xi_1_infinite_full_regions}. The bulk of the solutions start asymptotically at the unstable node $D_\pm$, and find themselves in one of the three types of regions between heteroclinic orbits lied out as follows: orbit $D_\pm \rightarrow F_\pm$, region $(I_\pm)$, orbit $D_\pm \rightarrow B_\pm$, region $(II_\pm)$, orbit $D_\pm \rightarrow B_\mp$, region $(III_\mp)$, orbit $D_\pm \rightarrow E_\mp$. Note that by the logic of the phase space the regions $(II_\pm)$ span both positive and negative zones of $\phi$ in a connected manner, although the heteroclinic orbits $D_\pm \rightarrow B_\mp$ are extremely close to the boundary of the unphysical region, and are not explicitly drawn on the plot. Perhaps this structure becomes more apparent when we turn of small $\Lambda$ and the system is not so singular at $\phi=0$, as shown on Fig.\ \ref{fig: quadratic_xi_1_infinite_full_regions_Lambda}. Taking $\Lambda$ ever smaller does not break the heteroclinic orbits and preserves the structure.

Solutions in the sectors described above have different fates. If a solution finds itself in the regions $(I_\pm)$ or $(III_\pm)$, it turns from deceleration to acceleration and ends up with eternal de Sitter type state at the points $F_\pm$ or $F_\mp$, respectively. Otherwise, if a solution finds itself in the regions $(II_\pm)$ it either just decelerates, or turns from deceleration to acceleration and deceleration again, ending up at late universe around the point $A$. (Since $A$ is a stable focus, the model would imply repeating oscillations between acceleration and deceleration in the late universe, but we may assume that reheating process and other types of matter will step in to influence the dynamics.) In the latter case, whether the era of transient acceleration is pronounced enough to produce 50-60 e-folds of expansion, depends on how soon a trajectory gets close to the master solution. That master orbits running from $B_\pm$ to $A$ attract all the trajectories in the regions $(II_\pm)$, but not all of those can reach its vicinity sufficiently early to experience enough accelerated expansion.

In the case of minimal coupling the global portraits are not so explicit, as the the whole asymptotics of $\phi$ is represented by the points $C_\pm$. In the limit $\xi\rightarrow 0$ the points $B_\pm$, $D_\pm$, $E_\pm$, and $F_\pm$ merge, which explains the strange hybrid ``saddle-node'' behavior around point $C_\pm$. However, the overall structure of the phase space is comparable to the sectors $(II_\pm)$ of the nonminimal case. There are heteroclinic orbits running out of $C_\pm$ into $A$ and only one class of generic solutions. These solutions either experience deceleration, or a sequence of deceleration, acceleration, and deceleration again, before settling into the oscillations of the post inflationary era.

\section{Quartic potential}
\label{sec: quartic potential}

As a second example let us consider quadratic nonminimal coupling and quartic potential,
\begin{equation}
\label{eq: quartic model}
F = 1 + \xi \phi^2 \,, \qquad V= \frac{\lambda}{4} \phi^4 + \Lambda \,.
\end{equation}
Like before, we take $\Lambda$ as a small regularizing parameter, and apply the limit $\Lambda \rightarrow 0$ to the presented equations and plots for $\xi=0, \, 1, \, 10, \, 10000$ on Fig. \ref{quartic_plots}. Let us mention that the notable Higgs inflation model \cite{Bezrukov:2007ep} (preceded by Refs.\ \cite{Spokoiny:1984bd,Barvinsky:1994hx}) where inflation is realized at the large values of $\phi$, corresponds to the parameters $\xi \approx 10^4, \lambda \approx 10^{-1}, \Lambda\approx 0$.

\subsection{Finite analysis}

\begin{figure*}
	\centering
	\subfigure[]{
		\includegraphics[width=5.5cm]{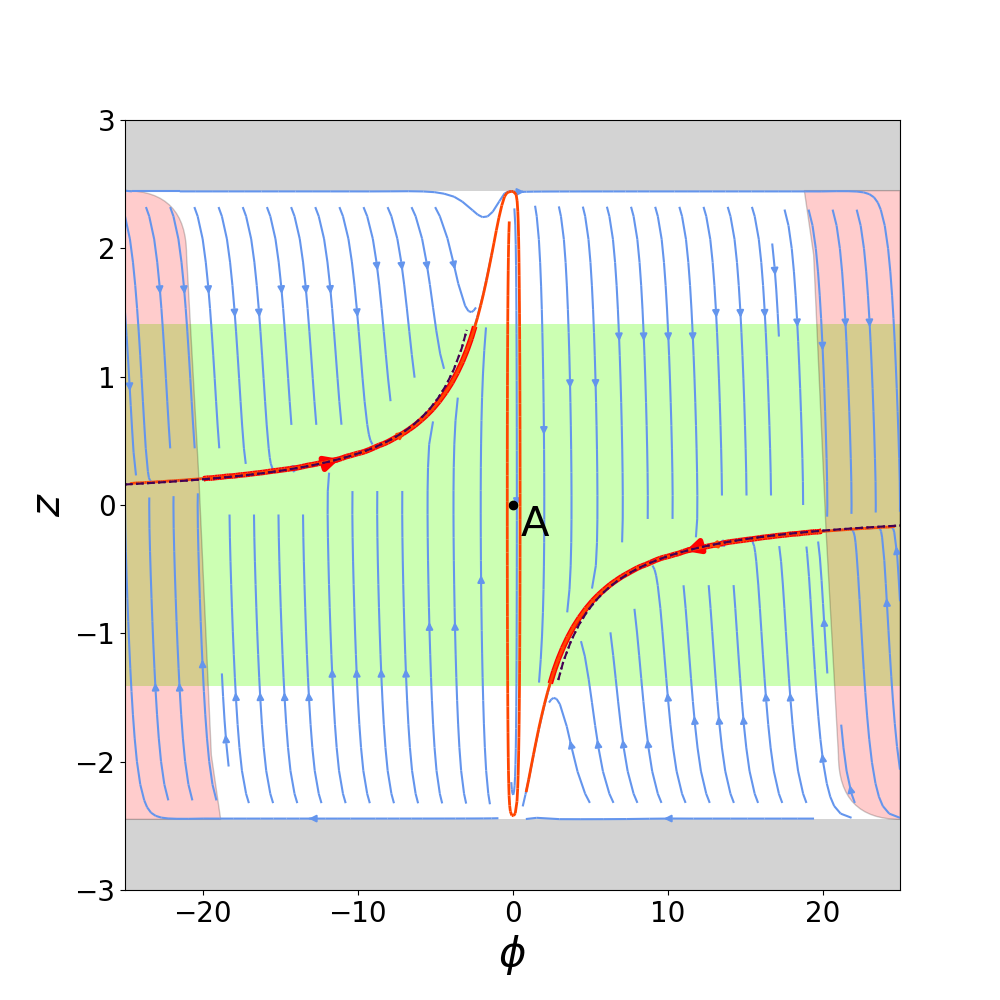} \label{fig: quartic_xi_0_finite}}
	\subfigure[]{
		\includegraphics[width=5.5cm]{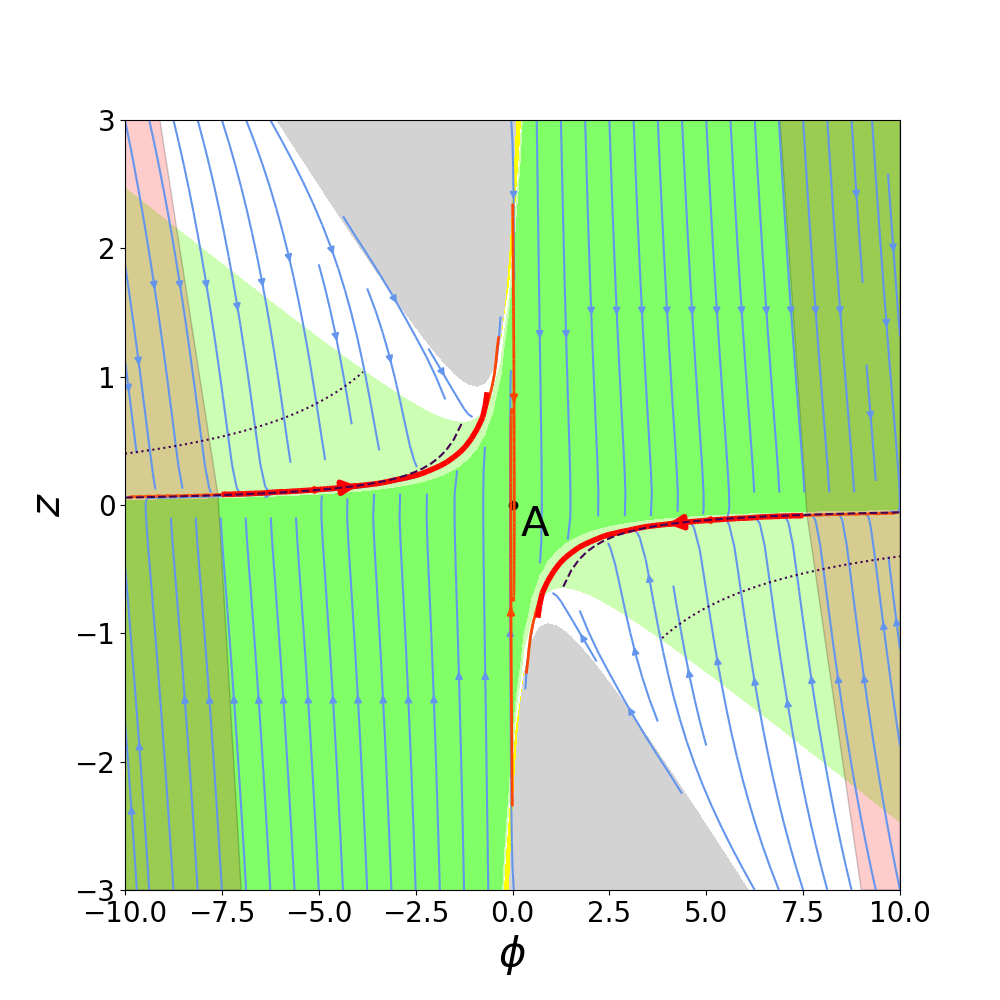} \label{fig: quartic_xi_1_finite}}
	\subfigure[]{
		\includegraphics[width=5.5cm]{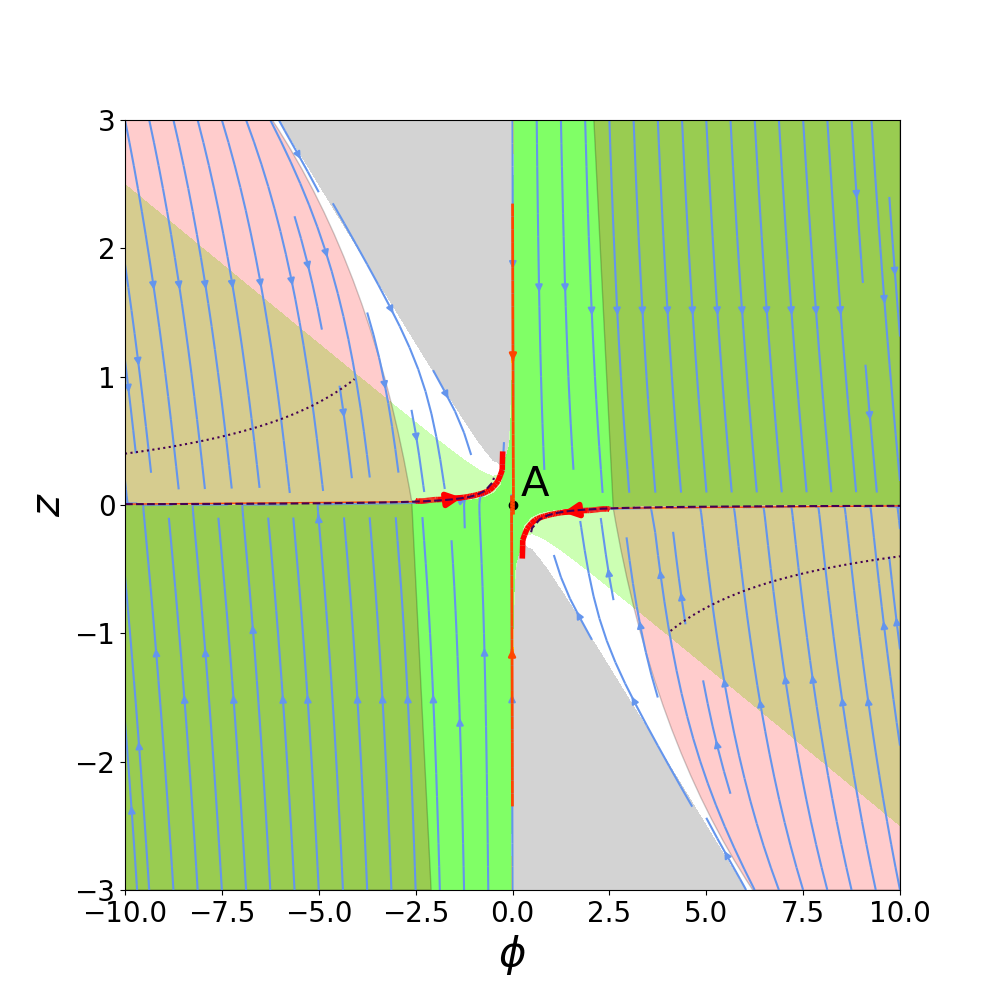} \label{fig: quartic_xi_10_finite}}
\\
	\subfigure[]{
		\includegraphics[width=5.5cm]{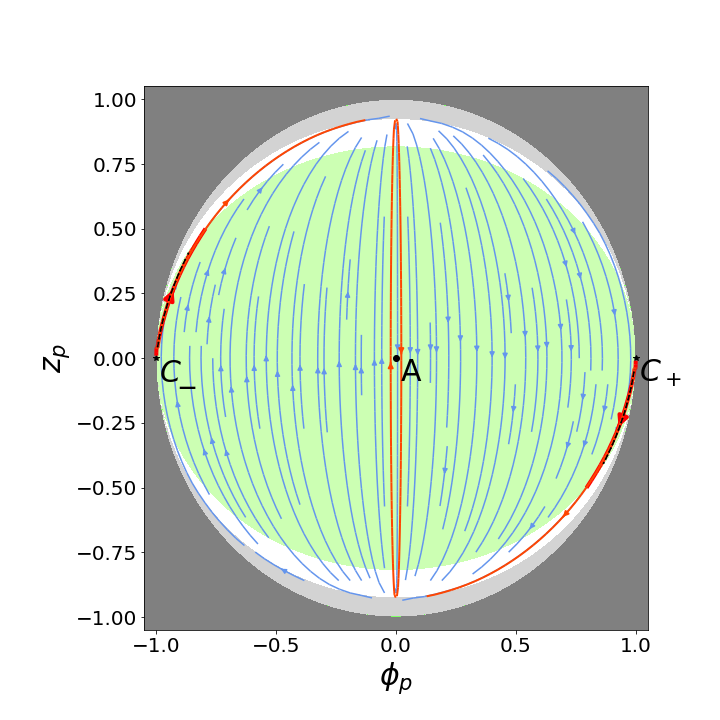} \label{fig: quartic_xi_0_infinite_full} }
	\subfigure[]{
		\includegraphics[width=5.5cm]{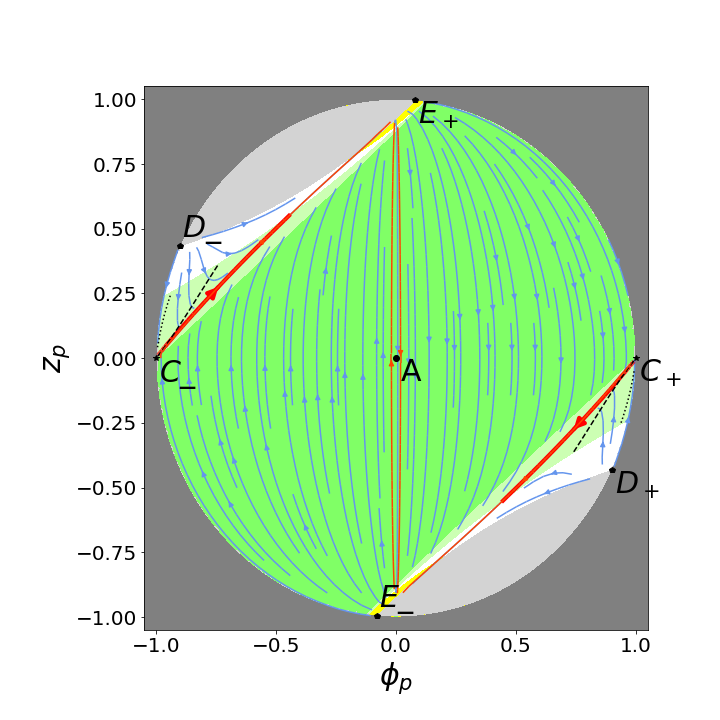} \label{fig: quartic_xi_1_infinite_full} }
	\subfigure[]{
		\includegraphics[width=5.5cm]{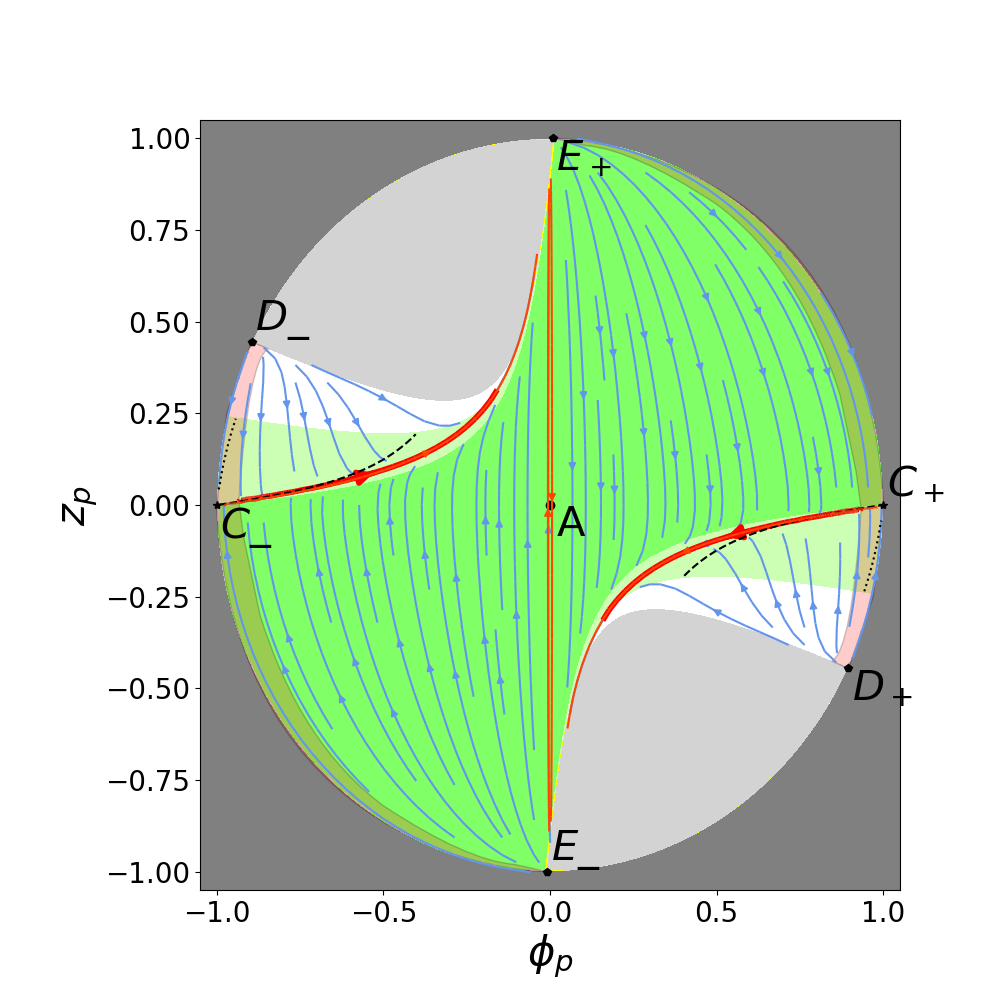} \label{fig: quartic_xi_10_infinite_full}}
\\
	\subfigure[]{
		\includegraphics[width=5.5cm]{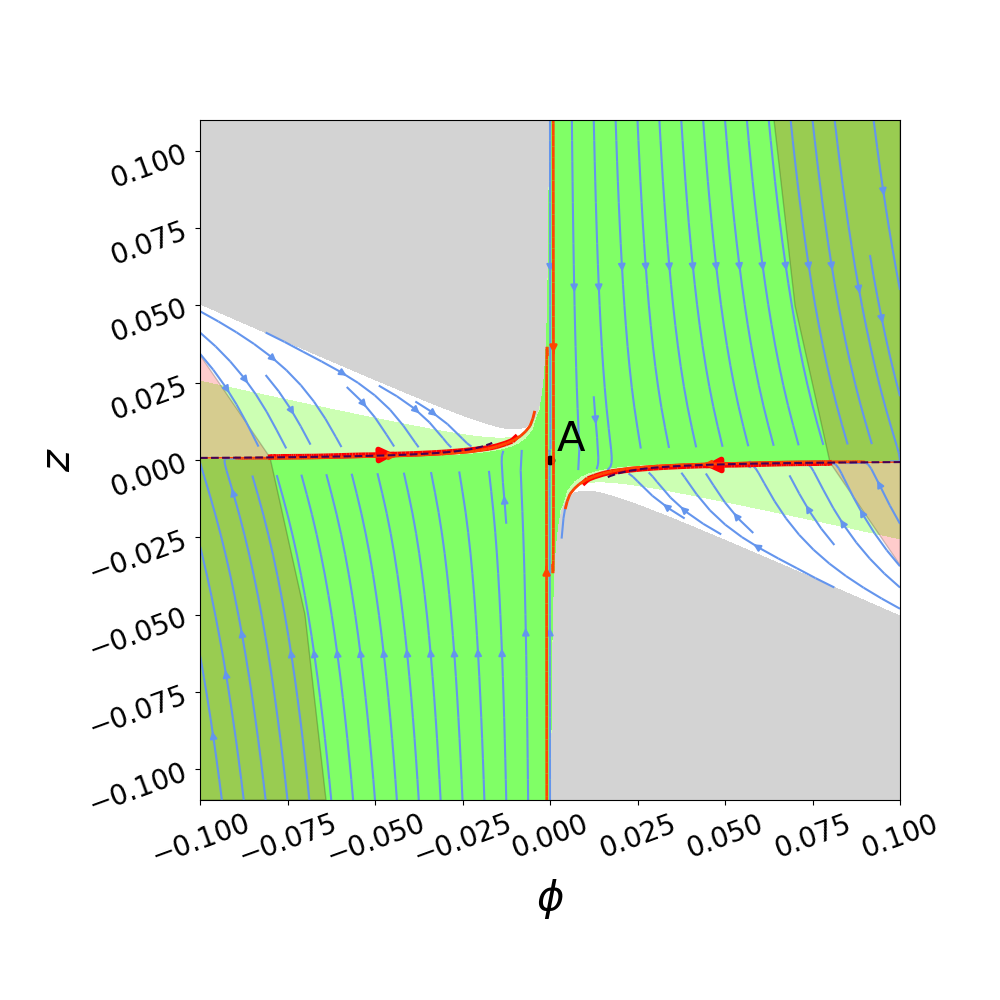}
    \label{fig: quartic_xi_10000_finite}}
	\subfigure[]{
		\includegraphics[width=5.5cm]{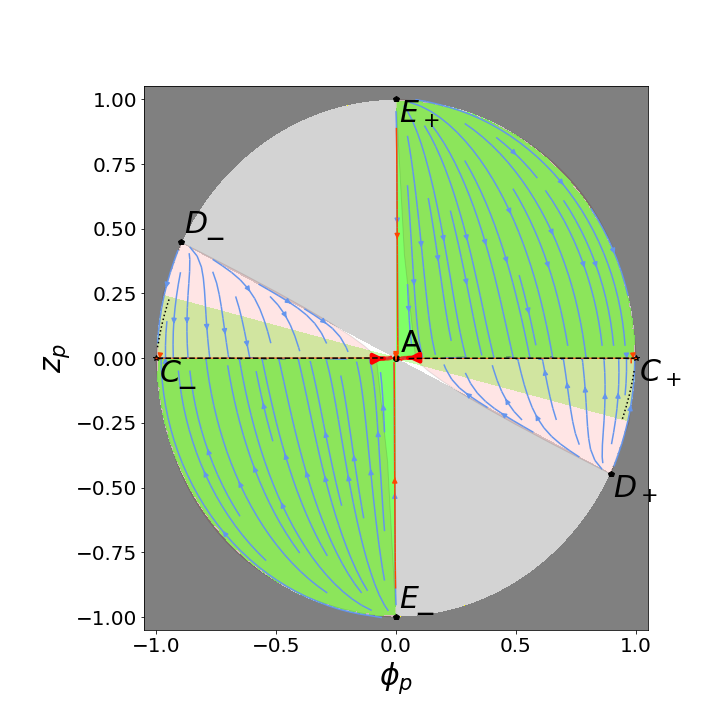}
    \label{fig: quartic_xi_10000_infinite_full}}
    \subfigure[]{
		\includegraphics[width=5.5cm]{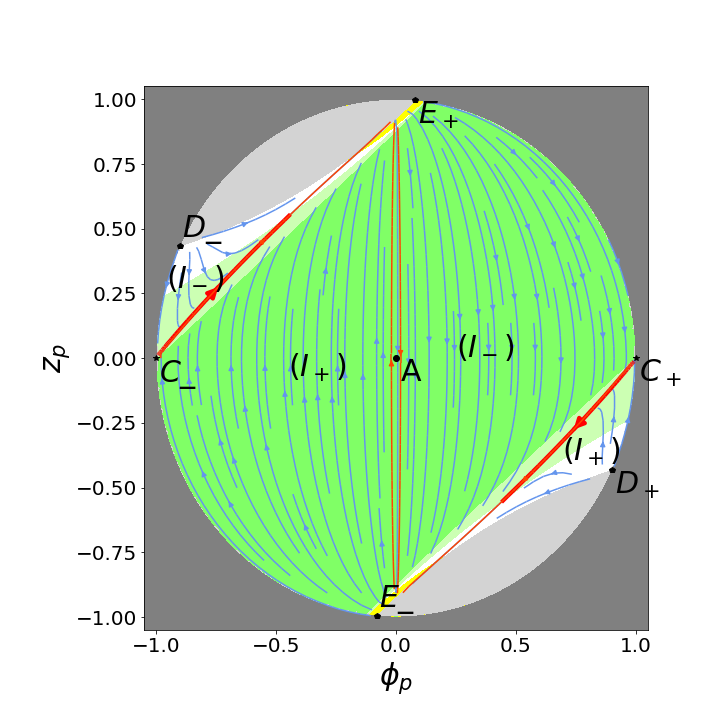}
    \label{fig: quartic_xi_1_infinite_full_regions}}
	\caption{Cosmological phase portraits of the quartic model \eqref{eq: quartic model} with $\Lambda=0$, and $\xi=0$ (panels a, d), $\xi=1$ (panels b, e, f), $\xi=10$ (panels c, f), $\xi=10000$ (panel g, h). Green background stands for superaccelerated, light green accelerated, white decelelerated, and yellow superstiff expansion, while grey covers the unphysical region. Orange trajectories are heteroclinic orbits between the fixed points, wider red part of the master trajectory highlights the 50 last e-folds of inflation, while red semi-transparent cover shows the basin of initial conditions leading to 50 e-folds. The dashed curve marks the path of slow roll approximation \eqref{eq: slow roll true z phi4}, while the dotted curve runs along the other slow roll approximation \eqref{eq: slow roll wrong z phi4} suggested in the literature.}
	\label{quartic_plots}
\end{figure*}

For the quartic model \eqref{eq: quartic model} the dynamical system \eqref{eq: dynsys phi}, \eqref{eq: dynsys z} reads
\begin{align}
\phi' &= z
\label{eq: dynsys phi phi4}\\
z' &= \frac{ (z^2 - 6) (\phi z + 4)}{2\phi (1 + \phi^2 \xi + 6\phi^2 \xi^2)} 
\nonumber \\ & \quad 
- \frac{\xi (3\phi^2 z + 5\phi z^2 + 12 \phi - 3 z^3 + 36 z)}{ (1 + \xi \phi^2 + 6 \xi^2 \phi^2)} 
\nonumber \\ & \quad 
- \frac{6 \xi^2 \phi z (3 \phi + 5z)  }{(1 + \xi \phi^2 + 6 \xi^2 \phi^2 )} \,,
\label{eq: dynsys z phi4}
\end{align}
and is independent of the coupling $\lambda$ in the potential.
Similarly, the barotropic index \eqref{eq: dynsys weff} 
\begin{align}
\weff &= -1 + \frac{z^2 + 2 \xi(3z^2-4 \phi z -12) -48 \xi^2 \phi z}{3 (1+ \xi \phi^2 + 6 \xi^2 \phi^2)} \,. 
\label{eq: dynsys weff phi4}
\end{align}
and the boundary of the physical phase space \eqref{eq: Friedmann constraint}, 
\begin{align}
z^2 - 6\xi \phi (\phi + 2z) < 6 \,,
\label{eq: dynsys boundary quartic}
\end{align}
do not depend on $\lambda$. Moreover, the bound \eqref{eq: dynsys boundary quartic} coincides with the bound \eqref{eq: dynsys boundary quadratic} in the quadratic potential case. The bound is exactly satisfied at the boundaries $z_b^\pm$ \eqref{eq: quadratic z_b} with effective barotropic index $\weff{}^\pm_b$ \eqref{eq: quadratic w_eff_b}, representing the kinetic dominance regime where to potential is not important. 
Although the dynamical systems equations \eqref{eq: dynsys phi phi2} are \eqref{eq: dynsys phi phi4} are different, one can still check explicitly that the solutions do not cross the boundary of Eq.\ \eqref{eq: dynsys boundary quartic}. The trajectories in the central area of the phase space are depicted on the top panels of Fig.\ \ref{quartic_plots}.

Formally there are three regular fixed points (attractor, saddle, attractor) but in the limit $\Lambda \rightarrow 0$ these merge into a single attractive fixed point at the origin $(0,0)$. This point is a stable node for $\xi\leq \tfrac{3}{32}$ and stable focus for $\xi>\tfrac{3}{32}$ \cite{Dutta:2020uha}. Since the dark energy scale important for the late universe is much lower than the inflation energy scale, distinguishing the precise structure of the late time attractor does not concern us much, as inflation will be well over before the trajectories reach there. What is important, however, is the slow roll curve \eqref{eq: slow roll true z} given by
\begin{align}
\label{eq: slow roll true z phi4}
z &= - \frac{4 (1+ \xi \phi^2)}{\phi (1 + \xi \phi^2 +  6\xi^2 \phi^2)} \,.
\end{align}
Note that it starts at the the scalar field asymptotics, $(\pm \infty, 0)$ and slowly diverges from $z=0$ as $|\phi|$ decreases. The asymptotic point corresponds to de Sitter like expansion ($\weff=-1$), but as we trace the slow roll curve towards the origin $\weff$ increases, until we leave the phase space region of accelerating expansion ($\weff<-\tfrac{1}{3}$). Beyond that it becomes meaningless to follow the slow roll curve any further, since inflation has ended and the premises of slow roll do not hold any more. Just for the record, the alternative slow roll curve \eqref{eq: slow roll wrong z} is now given by
\begin{align}
\label{eq: slow roll wrong z phi4}
z &=-\frac{4}{\phi} \,, 
\end{align}
and in comparison with \eqref{eq: slow roll true z phi4} it clearly fares much worse in approximating the true ``attractor solution'', see e.g.\ the plots \ref{fig: quartic_xi_1_finite}, \ref{fig: quartic_xi_10_finite}.
 Note, that in the limit $\phi \to \infty$ the ratio of the variables $z$ corresponding to these two slow roll proposals tends to constant value equal to $1+ 6 \xi$. Remembering that the alternative slow roll curve have been obtained
by setting $\dot H=0$ in the Klein-Gordon  equation, we can see that, despite the limit $\phi \to \infty$ clearly corresponds to $\dot H/H^2 \to 0$, the presence of $\dot H$ in the RHS of the Klein-Gordon equation
gives nevertheless a finite effect in scalar field dynamics. 

Contrary to the quadratic potential case, there are no regular saddle de Sitter fixed points in the interior of the phase space. However, the leading trajectory which collects the neighbouring trajectories into its vicinity and engenders inflationary expansion is still a heteroclinic orbit, running out from an asymptotic fixed point into $A$, as the global diagrams will reveal. On this trajectory we can identify the stretch of the path corresponding to the last 50 e-folds of expansion before the acceleration wanes, highlighed by extra red on the plots \ref{quartic_plots}. Although the trajectories will see some accelerated expansion before reaching into the close proximity of the ``attractor solution'', the contribution from the approaching ``fast roll'' stint amounts to a rather small value in terms of the e-folds $N$. 

The zone of good initial conditions which allow at least 50 e-folds of accelerated expansion compares well with the numerical results of Ref.\ \cite{Mishra:2019ymr}. The good zone stretches out from the asymptotics to a finite edge, which shifts closer to the origin as the nonminimal coupling $\xi$ increases. In this sense the quartic model behaves in the opposite way to the quadratic model. In the minimal coupling limit both models possess qualitatively similar basins of good initial conditions, but as $\xi$ increases, the good zone narrows down in the quadratic model, but expands in in the quartic model.

\subsection{Infinite analysis}

In the compact variables \eqref{eq: Poincare variables} the dynamical system \eqref{eq: dynsys phi phi4}, \eqref{eq: dynsys z phi4} translates into
\begin{align}
\phi_p' &= \frac{z_p (26 q^2 + 6 q \phi_p z_p - 2 q z_p^2 - \phi_p z_p^3)}{2 (q + \xi \phi_p^2 + 6 \xi^2 \phi_p^2)} 
\nonumber \\ & \quad 
+ \frac{\xi \phi_p z_p (13 q \phi_p + 36 q z_p + 3 \phi_p^2 z_p + 6 \phi_p z_p^2 - 3 z_p^3)}{q + \xi \phi_p^2 + 6 \xi^2 \phi_p^2} 
\nonumber \\ & \quad 
+ \frac{6 \xi^2 \phi_p^2 z_p (q + 3 \phi_p z_p + 6 z_p^2) }{q + \xi \phi_p^2 + 6 \xi^2 \phi_p^2} 
\end{align}
\begin{align}
z_p' &= -\frac{12 q^2( q + \phi_p^2)}{\phi_p (q + \xi \phi_p^2 + 6 \xi^2 \phi_p^2 )} 
\nonumber \\ & \quad 
-\frac{z_p (6 q^2 \phi_p + 6 q \phi_p^3 - 4 q^2 z_p - 2 q \phi_p^2 z_p - q \phi_p z_p^2 - \phi_p^3 z_p^2)}{2 \phi_p (q + \xi \phi_p^2 + 6 \xi^2 \phi_p^2 )} \nonumber \\ & \quad 
- \frac{12 \xi q \phi_p (q  + \phi_p^2)}{q+ \xi \phi_p2  + 6  \xi^2 \phi_p^2 } 
\nonumber \\ & \quad 
- \frac{\xi q z_p (36 q + 39 \phi_p^2 + 5 \phi_p z_p - 3 z_p^2)}{q+ \xi \phi_p2  + 6  \xi^2 \phi_p^2 } 
\nonumber \\ & \quad 
- \frac{3 \xi z_p (\phi_p^4 + 2 \phi_p^3 z_p -  \phi_p^2 z_p^2)}{q+ \xi \phi_p2  + 6  \xi^2 \phi_p^2 } 
\nonumber \\ & \quad 
- \frac{6 \xi^2 \phi_p z_p (3 q \phi_p + 5 q z_p + 3 \phi_p^3 + 6 \phi_p^2 z_p)}{q + \xi \phi_p^2 + 6  \xi^2 \phi_p^2 }
\end{align}
where we again use the shorthand notation $q=1-x^2-y^2$. 
The boundary of the physical phase space \eqref{eq: dynsys boundary quartic} is in the compact picture 
\begin{align}
\frac{z_p^2-6\xi \phi_p (\phi_p+2 z_p)}{q} < 6 \,.
\end{align}
Analogously, the barotropic index \eqref{eq: dynsys weff phi4} is given by
\begin{align}
\weff &= -1 + \frac{\phi_p^2+2\xi(12\phi_p^2 - 4 \phi_p z_p +15 z_p^2 -12) -48 \xi^2 \phi_p z_p}{3(q + \xi \phi_p^2 + 6 \xi^2 \phi_p^2)} \,,
\end{align}
the slow roll curve \eqref{eq: slow roll true z phi4} translates to an implicit form
\begin{align}
\label{eq: quartic slow roll phi_p z_p}
- \frac{4q^2 +q\phi_p z_p + \xi \phi_p^2(4q+\phi_p z_p)+6 \xi^2 \phi_p^3 z_p}{\sqrt{q}\phi_p (q + \xi \phi_p^2 + 6 \xi^2 \phi_p^2)} = 0 \,,
\end{align}
while the naive alternative slow roll curve \eqref{eq: slow roll wrong z phi4} maps to
\begin{align}
- \frac{4q +\phi_p z_p}{\phi_p\sqrt{q}} = 0 \,.
\end{align}
The finite attractive fixed point (in the $\Lambda \rightarrow 0$ limit) resides at
\begin{align}
    A & : (0, 0) \,
\end{align}
also in the compact variables. As sample of the corresponding global diagrams for different values of $\xi$ are presented on the lower panels of Fig.\ \ref{quartic_plots}.

Compactification of the phase space to a finite disc allows us to discern the features at the asymptotics, mapped to a unit circle. In the asymptotic limit $q \rightarrow 0$ the system reduces to
\begin{align}
\phi_p' &= - \frac{z_p^2 (z_p^2- 6\xi \phi_p( \phi_p +2 z_p))}{ 2 \xi \phi_p} \\
z_p' &= \frac{z_p (z_p^2- 6\xi \phi_p( \phi_p +2 z_p))}{ 2 \xi} \,,
\end{align}
where $z_p^2=1-\phi_p^2$.
For minimally as well as nonminimally coupled cases there are asymptotic fixed points 
\begin{align}
C_\pm &: (\pm 1, 0) \,,
\end{align}
whereby $\weff=-1$. Discussion in the next section \ref{sec: asymptotic regimes} allows us to identify the actual regime these points correspond to, purporting the name asymptotic de Sitter. 
The eigenvalues are $-3$ and $0$, so the points are nonhyperbolic. The attractive eigendirection lies tangent to the outer boundary of the phase space, while the eigendirection corresponding to the zero eigenvalue is $(\frac{1+6\xi}{8}, \, 1)^T$. The latter direction launches an heteroclinic orbit that reaches the central fixed point $A$. This orbit in none other but the ``attractor solution'' to which other trajectories converge and which provides the principal path for inflation. Since the eigendirection of the zero eigenvalue is tangent to the slow roll curve \eqref{eq: quartic slow roll phi_p z_p} at this point we feel justified to trust the curve \eqref{eq: quartic slow roll phi_p z_p} as a meaningful first approximation for the master trajectory.

In the case of nonminimal coupling, there are also fixed points where the boundary of the physical phase space reaches the asymptotic circle:
\begin{align}
D_\pm &: (\pm\Omega_+,\mp \sqrt{1-\Omega^2_+}) \\
E_\pm &: (\pm\Omega_-,\pm \sqrt{1-\Omega^2_-}) 
\end{align}
where $\Omega_\pm$ is given by Eq.\ \eqref{eq: Omega_pm}. Recall that the boundary of the physical phase space depended only on the nonminimal coupling, and not the protential, so these points are analogous to the quadratic case. The points $D$ have expansion corresponding the barotropic index $\weff{}_{,b}^-$ while the points $E$ have barotropic index $\weff{}_{,b}^+$, given in Eq.\ \eqref{eq: quadratic w_eff_b}. We will discuss them further in Sec.\ \ref{sec: asymptotic regimes}.

The overall structure of the phase space is very simple, as the heteroclinic orbits flowing out from the fixed point $C_\pm$ to the fixed point $A$ divide the chart into two regions. In the nonminimal coupling case on Fig.\ \ref{fig: quadratic_xi_1_infinite_full_regions} generic trajectories start at the point $D_\pm$ and either experience phases of deceleration, acceleration, and deceleration, or just deceleration, before the early universe era ends, the oscillations add new physics and the model reaches its limit of applicability. One can see perhaps more explicitly how the regions are connected by turning on a small $\Lambda$ like on Fig.\ \ref{fig: quartic_xi_1_infinite_full_regions_Lambda}. (As mentioned before, the late time dynamics with $\Lambda>0$ is slightly modified by the splitting of the point $A$ into three separate points, however the features describing early universe remain intact.)

In the minimal coupling limit the asymptotic fixed points $C_\pm$, $D_\pm$, $E_\pm$ merge, and the remaining $C_\pm$ inherits strange saddle-node characteristics. Inflation is still ruled by the heteroclinic orbits from $C_\pm$ to $A$. Generic trajectories start near the points $C_\pm$ and are attracted by the heteroclinic orbit, but only those trajectories which get close to the ``attractor solution'' soon enough will experience sufficient number e-folds in accelerated expansion.

\section{Asymptotic regimes}
\label{sec: asymptotic regimes}

The set of variables we use in the present paper have their advantages in the graphical representation of the cosmological dynamics, however, from the physical point of view, it is important to express all asymptotic regimes found in terms of the original variables. In this section we can actually take a shortcut: since the asymptotic regimes in the $(\phi, \dot \phi)$ variables for nonminimally coupled scalar field are already listed in the literature, we first recall the explicit forms of the known asymptotic regimes for the models under investigation, identify them with the stable points in our phase diagrams, and confirm that none of the known regimes has been projected to the same point in a diagram. Hence all regimes can be distinguished on the phase diagrams clearly.

We start with the asymptotics of the minimally coupled scalar field, studied in Refs.\ \cite{Felder:2002jk,Foster:1998zu,Foster:1998sk,Leon:2008de,Handley:2014bqa}, and use it to illustrate the problem of fixed point projections.
The first regime to mention is the kinetic dominant regime. It is known that this regime is an attractor for a minimally coupled scalar field with a potential less steep than an exponential one, at the stage of Universe contraction, thus it is an unstable node for an expanding Universe. In this regime the potential energy of a scalar field is negligible with respect to the kinetic part, and the field is effectively massless. The explicit form of this  asymptotic is $a\sim t^{1/3}$, $|\phi| \sim \sqrt{\tfrac{2}{3}} \ln{t}$, so that $H\sim \tfrac{1}{3t}$ and $|\dot \phi| \sim \tfrac{\sqrt{2}}{\sqrt{3}t}$. This means that $|\phi| \to \infty$ and our variable $z$ tends to a constant value  $|z| \to \sqrt{6}$. After Poincar\'e compactification we evidently have $|\phi_p| \to 1$ and $z_p \to 0$.
The second important regime for a minimally coupled scalar field  is the well-known potential dominated inflation. The inflationary asymptotic for a massive scalar field gives $|\dot \phi| = const$ while $H\sim |\phi|$ which means that in the infinite past $|\phi| \to \infty$ and $z \to 0$. Hence after the Poincar\'e compactification this regime is mapped to the same points ($\pm 1, 0$) as the kinetic regime. Analogously for the quartic potential where $|\phi| \sim e^{-t}$ and $H \sim \phi^2$ the asymptotic behavior also leads to $z \to 0$. 
This means that despite these two regimes are clearly distinguishable on the $(\phi, z)$ diagram (see  Figs.\ \ref{fig: quadratic_xi_0_finite} and \ref{fig: quartic_xi_0_finite} where the kinetic regime starts from the boundary line $|z|=\sqrt{6}$, while the orange curve that represents inflation starts from $z=0$), they are projected onto the same point in the $(\phi_p, z_p)$ plane ($C_+$ or $C_-$ depending on the sign of $\phi$) which makes the corresponding global phase diagrams on Figs.\ \ref{fig: quadratic_xi_0_infinite_full} and \ref{fig: quartic_xi_0_infinite_full} more difficult to interpret.

Before turning to particular regimes in cosmology with a nonminimally coupled scalar field let's make a general statement about power-law regimes. They are very typical in cosmology and we can see an advantage to use the $(\phi, z)$ coordinates instead of direct $(\phi, \dot \phi)$ to distinguish them. Indeed, let $a \sim t^{\alpha}$ and $\phi \sim t^{\beta}$. Different $\alpha$ gives us different effective equations of state since evidently $\weff= \tfrac{2}{3 \alpha} - 1$.
However, as $\dot \phi \sim t^{\beta -1}$, the ratio $\tfrac{\dot \phi}{\phi}$ tends to infinity at $t \to 0$, and if we introduce
compactified variables in analogy with Eq.\ \eqref{eq: Poincare variables},
\begin{align}
\phi_p = \frac{\phi}{\sqrt{1+\phi^2+\dot \phi^2}}\,, \qquad
\dot \phi_p = \frac{\dot \phi}{\sqrt{1+\phi^2+\dot \phi^2}} \,
\end{align}
we get that in compactified coordinates $\phi_p$ and $\dot \phi_p$  {\it any} power-law singular solution maps into the particular point $\phi_p=0, \dot \phi_p=1$. 
On the contrary, dividing $\dot \phi$ by $H$ we compensate for this additional power since the power-law dependence of the scale factor implies $H \sim \tfrac{1}{t}$. This means that $z$ is proportional to $\phi$ with the coefficient $\kappa = \tfrac{\beta}{\alpha}$ in our notation. This easily leads to the following coordinates of the corresponding compactified variables 
\begin{align}
\phi_p &= \frac{1}{\sqrt{1+\kappa^2}} \,, \qquad
z_p &= \frac{\kappa}{\sqrt{1+\kappa^2}} \,.
\end{align}
Therefore different power-law regimes can be clearly distinguished in $(\phi, z)$. 

In the non-minimal coupling case for both quadratic and quartic potentials instead of a single kinetic dominant regime there are two power-law asymptotics   \cite{Carloni:2007eu}, an unstable node with 
\begin{align}
\alpha &= \frac{1}{3+12\xi - 2\sqrt{6\xi(6\xi+1)}} \,, \nonumber \\  
\beta &= (6\xi - \sqrt{6\xi(6\xi+1)}) \alpha \,,
\end{align}
and a saddle with
\begin{align}
\alpha& = \frac{1}{3+12\xi+2\sqrt{6\xi(6\xi+1)}} \,, \nonumber \\ 
\beta&=(6\xi+\sqrt{6\xi(6\xi+1)}) \alpha \,.
\end{align}
Evidently, the coordinates $\phi_p$ for these two points
can be written as a single expression
\begin{equation}
\label{eq: phi_p carloni}
    \phi_p=\frac{1}{\sqrt{1+6\xi+72\xi^2 \pm 12\xi\sqrt{6\xi+36\xi^2}}},
\end{equation}
where the $+$ sign is for the saddle, and the $-$ sign is for the node. A closer inspection revels that the expression \eqref{eq: phi_p carloni} coincides with our result \eqref{eq: Omega_pm}, which establishes that the node is our point $D$, and the saddle is our point $E$. Computing the $\weff$ from $\alpha$ reproduces the expression \eqref{eq: quadratic w_eff_b} as well.

In the case of a massive scalar field there is a large field attractor regime $a \sim e^{Ht}$,
$|\phi| \sim e^{-\tfrac{2H \xi t}{1+4 \xi}}$   \cite{Carloni:2007eu,Sami:2012uh}. It is realized when the the value of the scalar field is bigger than at the unstable de Sitter point, so that instead of rolling towards $|\phi| \to 0$ it runs away to infinity.
In this regime $H=const$, so that both $\phi$ and $z$ tend to infinity exponentially in time. This means that $\phi_p$ is proportional to $z_p$ and the fixed point is located  between $(0,\pm 1)$ and $(\pm 1,0)$ at the limiting circle. Using the explicit form of the exponential solution we get that the coordinate $\phi_p$ is equal to
\begin{equation}
|\phi_p| =\frac{1+4\xi}
{\sqrt{1 + 8 \xi + 20 \xi^2}}.
\end{equation}
This expression coincides with the expression \eqref{eq: Omega_F} and indicates that the corresponding regime is represented by the point $F$ in our variables. Although formally we have a de Sitter situation with constant $H$, the scalar field keeps advancing up the potential while the effective gravitational constant decreases in time.
The massive case also possesses an unstable de Sitter solution with $\dot \phi=0$, it is evidently  mapped into the points $B\pm$. The inflationary trajectory should have its origin near the point $B_\pm$. Thus inflation for the case of a quadratic potential can not be realized for large enough $|\phi|$ in sharp contrast with the case of the 
minimally coupled field.

In the case of quartic potential there are no unstable de Sitter points. But there is an inflationary regime with arbitrary large scalar field with the Hubble parameter $H$ and the scalar field $\phi$ growing linearly in time when $t$ is directed to minus infinity: $H \sim t$, $|\phi| \sim t$ \cite{Skugoreva:2014gka}. This form means that $\dot \phi$ tends to a constant and thus $z \to 0$. So, the corresponding fixed point is $|\phi_p| =1$, $z_p=0$. We can recognize our points $C_\pm$. It should be noted that despite the effective barotropic index $\weff$ shows a typical de Sitter value of $-1$ in this regime, the Hubble function evolves in time, and the overall situation is rather different from the usual de Sitter. It makes sense to use another name like `asymptotic de Sitter' to distinguish this case from the regular de Sitter.

We see that for nonminimal coupling not only different power-law regimes are mapped onto different fixed points at the compactified plane, but other known regimes also represent different fixed points. This makes that particular set of variables useful for graphic representation of the underlying dynamics. We note that this general property does not follow from some underlying principle (unlike the situation when different power-law regimes map onto different points, which can be shown explicitly for any power-law regime), and is a matter of (fortunate) coincidence. An opposite example has been
presented above for a minimally coupled scalar field when the kinetic and inflationary regimes accidentally map into
a single point on the global diagram. This does not happen for $\xi \ne 0$ since in this case the kinetic regime is of power-law nature, while for $\xi=0$ scalar field in the kinetic regime evolves logarithmically.

\section{Conclusions}
\label{sec: conclusions}

In the present paper we revisit the construction of phase diagrams for the cosmological dynamics with a scalar field nonminimally coupled to gravity, focusing upon the Jordan frame. In contrast to previous works we employ the coordinates $(\phi, \tfrac{\dot \phi}{H})$. This seemingly asymmetric form (as the ``speed'' of the scalar field is divided by $H$ while the value of the scalar field itself is not) has one important advantage. Namely, after introducing compactified variables by the standard \\
Poincar\'{e} procedure \eqref{eq: Poincare variables} this set of variables can clearly separate different asymptotic power-law regimes. This feature can be contrasted with the compactified version of the original variables $(\phi, \dot \phi)$, whereby every power-law regime would correspond to the same point on the global phase diagram, making it more difficult to interpret. 
It appears that the advantage of the $(\phi, \tfrac{\dot \phi}{H})$ set is not restricted to just power-law regimes  -- we show that the known inflationary points are not superimposed on any of the power-law points, and can be also clearly distinguished on the phase portraits. The obvious disadvantage of the used set in comparison with the original $(\phi, \dot \phi)$ variables is that it can not reliably describe the oscillations of a scalar field around the minimum of the potential in a nonsingular way if the potential has zero value at the minimum. However, as long as the goal of a study is the description of the regimes with high $|\phi|$, this disadvantage can
be regarded as a reasonable ``price'' for not mixing different $\phi$ asymptotics on the phase diagrams.

Our investigations stress and illustrate how physically relevant inflationary regimes can not be considered just as particular fixed points on a phase diagram (this would imply eternal inflation) but are rather represented by heteroclinic trajectories between the fixed points. We show that inflation does not have to be realized by only a center submanifold originating from a nonhyperbolic asymptotic fixed point \cite{Alho:2014fha}, but it can also happen with an unstable submanifold originating from a regular saddle point. Although the points where the inflationary orbit starts exhibit de Sitter like effective barotropic index, we are careful to distinguish between true de Sitter and asymptotic de Sitter points, regarding the actual physical situation there.

The heteroclinic orbits mentioned above are none other than the ``attractor solutions'' to which the neighbouring trajectories converge and which enjoy accelerated de Sitter like expansion due to the slowly varying scalar field. While the precise analytic form of these master trajectories is hard to pin down, they can be approximated by curves that can be derived from the slow roll conditions. Indeed, by manipulating the scalar field equation we give a simple analytic form of the approximate slow roll curve in the Jordan frame. In fact, this curve coincides with the slow roll condition obtained in the Einstein frame and translated into the Jordan frame. On the phase portraits we can witness that this curve is providing a better approximation to the full nonlinear ``attractor solution'' than the generalized slow roll proposal given in the literature.

We supply the phase diagrams by an extra color indication of the zone of initial conditions which lead to the physically required 50 e-folds of accelerated expansion. Since nonminimal coupling makes the effective Planck mass to depend on the scalar field, we do not restrict our attention only to the range of initial values that would correspond to Planck density in the units of the late universe, and bring into view full trajectories not just the initial data (which can correspond to the same trajectory). The diagrams show how turning on the nonminimal coupling $\xi$ affects the extent of the basin of good initial conditions. For quartic potentials the range of initial conditions that lead to at least 50 e-folds of expansion gets larger with increasing $\xi$, but for quadratic potentials this range shrinks down becoming almost negligible already at $\xi=1$.

It would be interesting to test these methods and insights upon other inflationary models which are in abundance in the literature \cite{Martin:2013tda}, including a scalar nonminimally coupled to curvature in the Palatini formulation \cite{Bauer:2008zj,Tenkanen:2020dge} (e.g.\ asking how are the phase spaces related in metric and Palatini models which otherwise give identical predictions \cite{Jarv:2020qqm}), but also in the teleparallel setting where the scalar is nonminimally coupled to torsion \cite{Geng:2011aj,Hohmann:2018rwf} or nonmetricity \cite{Jarv:2018bgs} instead of curvature. Understanding well the phase space features of successful inflationary models might also help to  construct new models with good properties.

\subsection*{Acknowledgements}
LJ was supported by the Estonian Research Council grant PRG356 and by the EU through the European Regional Development Fund CoE program TK133 “The Dark Side of the Universe.” AT is supported by the Russian Government Program of Competitive Growth of Kazan Federal University and RSF grant 21-12-00130.

\appendix
\section{Nonvanishing cosmological constant}
\label{sec: appendix}
\begin{figure*}[h!]
	\centering
	\subfigure[]{
		\includegraphics[width=5.5cm]{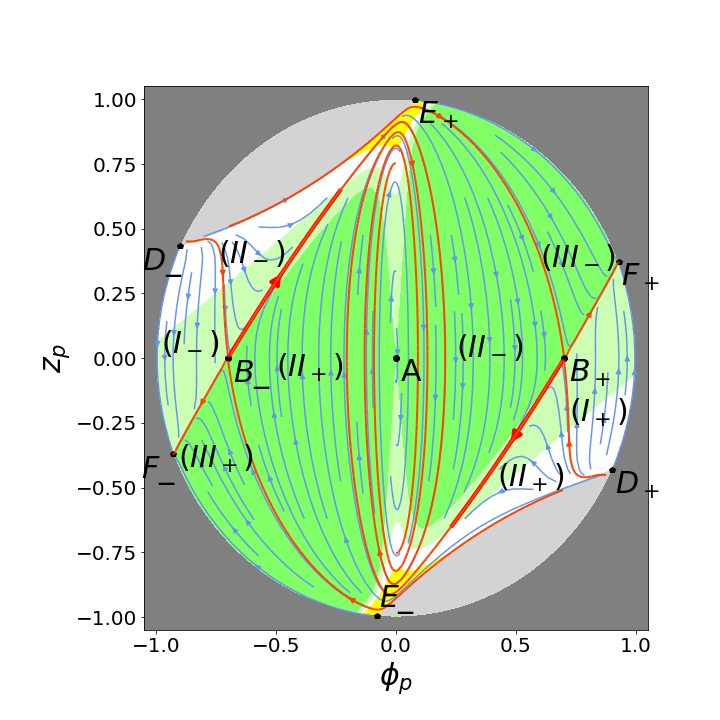} 
		\label{fig: quadratic_xi_1_infinite_full_regions_Lambda} }
	\subfigure[]{
		\includegraphics[width=5.5cm]{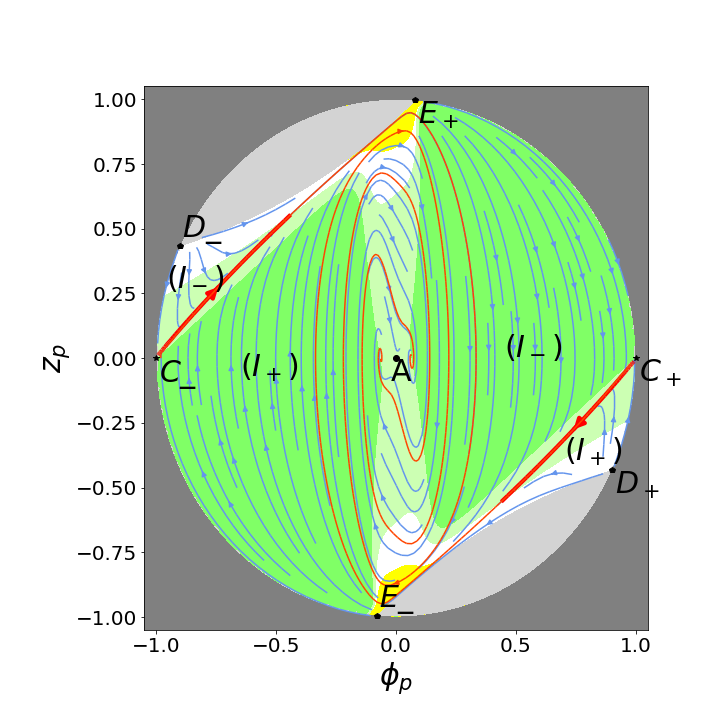} 
		\label{fig: quartic_xi_1_infinite_full_regions_Lambda} }
	\caption{Cosmological phase portrait of the quadratic model \eqref{eq: quadratic model} with $\xi=1$, $m=1$, $\Lambda=0.01$ (panel a) and quartic model \eqref{eq: quartic model} with $\xi=1$, $\lambda=1$, $\Lambda=0.001$ (panel b). Green background stands for superaccelerated, light green accelerated, white decelelerated, and yellow superstiff expansion, while grey covers the unphysical region. Orange trajectories are heteroclinic orbits between the fixed points, wider red part of the master trajectory highlights the 50 last e-folds of inflation.}
	\label{Lambda_plots}
\end{figure*}

Perhaps a more clear intuition of the regions of the global phase portrait can be obtained when a small constant term in the potential is turned on, as depicted on Fig.\ \ref{Lambda_plots}. This should be compared with the plots \ref{fig: quadratic_xi_1_infinite_full_regions} and \ref{fig: quartic_xi_1_infinite_full_regions}.

\bibliographystyle{utphys.bst}    

\providecommand{\href}[2]{#2}\begingroup\raggedright
\endgroup

\end{document}